
\catcode `\@=11 

\def\@version{1.3}
\def\@verdate{28.11.1992}


%
%
%
%
%
%

\font\fiverm=cmr5
\font\fivei=cmmi5	\skewchar\fivei='177
\font\fivesy=cmsy5	\skewchar\fivesy='60
\font\fivebf=cmbx5

\font\sevenrm=cmr7
\font\seveni=cmmi7	\skewchar\seveni='177
\font\sevensy=cmsy7	\skewchar\sevensy='60
\font\sevenbf=cmbx7

\font\eightrm=cmr8
\font\eightbf=cmbx8
\font\eightit=cmti8
\font\eighti=cmmi8			\skewchar\eighti='177
\font\eightmib=cmmib10 at 8pt	\skewchar\eightmib='177
\font\eightsy=cmsy8			\skewchar\eightsy='60
\font\eightsyb=cmbsy10 at 8pt	\skewchar\eightsyb='60
\font\eightsl=cmsl8
\font\eighttt=cmtt8			\hyphenchar\eighttt=-1
\font\eightcsc=cmcsc10 at 8pt
\font\eightsf=cmss8

\font\ninerm=cmr9
\font\ninebf=cmbx9
\font\nineit=cmti9
\font\ninei=cmmi9			\skewchar\ninei='177
\font\ninemib=cmmib10 at 9pt	\skewchar\ninemib='177
\font\ninesy=cmsy9			\skewchar\ninesy='60
\font\ninesyb=cmbsy10 at 9pt	\skewchar\ninesyb='60
\font\ninesl=cmsl9
\font\ninett=cmtt9			\hyphenchar\ninett=-1
\font\ninecsc=cmcsc10 at 9pt
\font\ninesf=cmss9

\font\tenrm=cmr10
\font\tenbf=cmbx10
\font\tenit=cmti10
\font\teni=cmmi10		\skewchar\teni='177
\font\tenmib=cmmib10	\skewchar\tenmib='177
\font\tensy=cmsy10		\skewchar\tensy='60
\font\tensyb=cmbsy10	\skewchar\tensyb='60
\font\tenex=cmex10
\font\tensl=cmsl10
\font\tentt=cmtt10		\hyphenchar\tentt=-1
\font\tencsc=cmcsc10
\font\tensf=cmss10

\font\elevenrm=cmr10 scaled \magstephalf
\font\elevenbf=cmbx10 scaled \magstephalf
\font\elevenit=cmti10 scaled \magstephalf
\font\eleveni=cmmi10 scaled \magstephalf	\skewchar\eleveni='177
\font\elevenmib=cmmib10 scaled \magstephalf	\skewchar\elevenmib='177
\font\elevensy=cmsy10 scaled \magstephalf	\skewchar\elevensy='60
\font\elevensyb=cmbsy10 scaled \magstephalf	\skewchar\elevensyb='60
\font\elevensl=cmsl10 scaled \magstephalf
\font\eleventt=cmtt10 scaled \magstephalf	\hyphenchar\eleventt=-1
\font\elevencsc=cmcsc10 scaled \magstephalf
\font\elevensf=cmss10 scaled \magstephalf

\font\fourteenrm=cmr10 scaled \magstep2
\font\fourteenbf=cmbx10 scaled \magstep2
\font\fourteenit=cmti10 scaled \magstep2
\font\fourteeni=cmmi10 scaled \magstep2		\skewchar\fourteeni='177
\font\fourteenmib=cmmib10 scaled \magstep2	\skewchar\fourteenmib='177
\font\fourteensy=cmsy10 scaled \magstep2	\skewchar\fourteensy='60
\font\fourteensyb=cmbsy10 scaled \magstep2	\skewchar\fourteensyb='60
\font\fourteensl=cmsl10 scaled \magstep2
\font\fourteentt=cmtt10 scaled \magstep2	\hyphenchar\fourteentt=-1
\font\fourteencsc=cmcsc10 scaled \magstep2
\font\fourteensf=cmss10 scaled \magstep2

\font\seventeenrm=cmr10 scaled \magstep3
\font\seventeenbf=cmbx10 scaled \magstep3
\font\seventeenit=cmti10 scaled \magstep3
\font\seventeeni=cmmi10 scaled \magstep3	\skewchar\seventeeni='177
\font\seventeenmib=cmmib10 scaled \magstep3	\skewchar\seventeenmib='177
\font\seventeensy=cmsy10 scaled \magstep3	\skewchar\seventeensy='60
\font\seventeensyb=cmbsy10 scaled \magstep3	\skewchar\seventeensyb='60
\font\seventeensl=cmsl10 scaled \magstep3
\font\seventeentt=cmtt10 scaled \magstep3	\hyphenchar\seventeentt=-1
\font\seventeencsc=cmcsc10 scaled \magstep3
\font\seventeensf=cmss10 scaled \magstep3

\def\@typeface{Computer Modern} 

\def\hexnumber@#1{\ifnum#1<10 \number#1\else
 \ifnum#1=10 A\else\ifnum#1=11 B\else\ifnum#1=12 C\else
 \ifnum#1=13 D\else\ifnum#1=14 E\else\ifnum#1=15 F\fi\fi\fi\fi\fi\fi\fi}

\def\mib{\hexnumber@\mibfam}
\def\syb{\hexnumber@\sybfam}

\def\makestrut{%
  \setbox\strutbox=\hbox{%
    \vrule height.7\baselineskip depth.3\baselineskip width 0pt}%
}

\def\bls#1{%
  \normalbaselineskip=#1%
  \normalbaselines%
  \makestrut%
}

%

\newfam\mibfam 
\newfam\sybfam 
\newfam\scfam  
\newfam\sffam  

\def\em{\ifdim\fontdimen1\font>0 \rm\else\it\fi}

\textfont3=\tenex
\scriptfont3=\tenex
\scriptscriptfont3=\tenex

\def\eightpoint{
  \def\rm{\fam0\eightrm}%
  \textfont0=\eightrm \scriptfont0=\sevenrm \scriptscriptfont0=\fiverm%
  \textfont1=\eighti  \scriptfont1=\seveni  \scriptscriptfont1=\fivei%
  \textfont2=\eightsy \scriptfont2=\sevensy \scriptscriptfont2=\fivesy%
  \textfont\itfam=\eightit\def\it{\fam\itfam\eightit}%
  \textfont\bffam=\eightbf%
    \scriptfont\bffam=\sevenbf%
      \scriptscriptfont\bffam=\fivebf%
  \def\bf{\fam\bffam\eightbf}%
  \textfont\slfam=\eightsl\def\sl{\fam\slfam\eightsl}%
  \textfont\ttfam=\eighttt\def\tt{\fam\ttfam\eighttt}%
  \textfont\scfam=\eightcsc\def\sc{\fam\scfam\eightcsc}%
  \textfont\sffam=\eightsf\def\sf{\fam\sffam\eightsf}%
  \textfont\mibfam=\eightmib%
  \textfont\sybfam=\eightsyb%
  \bls{10pt}%
}

\def\ninepoint{
  \def\rm{\fam0\ninerm}%
  \textfont0=\ninerm \scriptfont0=\sevenrm \scriptscriptfont0=\fiverm%
  \textfont1=\ninei  \scriptfont1=\seveni  \scriptscriptfont1=\fivei%
  \textfont2=\ninesy \scriptfont2=\sevensy \scriptscriptfont2=\fivesy%
  \textfont\itfam=\nineit\def\it{\fam\itfam\nineit}%
  \textfont\bffam=\ninebf%
    \scriptfont\bffam=\sevenbf%
      \scriptscriptfont\bffam=\fivebf%
  \def\bf{\fam\bffam\ninebf}%
  \textfont\slfam=\ninesl\def\sl{\fam\slfam\ninesl}%
  \textfont\ttfam=\ninett\def\tt{\fam\ttfam\ninett}%
  \textfont\scfam=\ninecsc\def\sc{\fam\scfam\ninecsc}%
  \textfont\sffam=\ninesf\def\sf{\fam\sffam\ninesf}%
  \textfont\mibfam=\ninemib%
  \textfont\sybfam=\ninesyb%
  \bls{12pt}%
}

\def\tenpoint{
  \def\rm{\fam0\tenrm}%
  \textfont0=\tenrm \scriptfont0=\sevenrm \scriptscriptfont0=\fiverm%
  \textfont1=\teni  \scriptfont1=\seveni  \scriptscriptfont1=\fivei%
  \textfont2=\tensy \scriptfont2=\sevensy \scriptscriptfont2=\fivesy%
  \textfont\itfam=\tenit\def\it{\fam\itfam\tenit}%
  \textfont\bffam=\tenbf%
    \scriptfont\bffam=\sevenbf%
      \scriptscriptfont\bffam=\fivebf%
  \def\bf{\fam\bffam\tenbf}%
  \textfont\slfam=\tensl\def\sl{\fam\slfam\tensl}%
  \textfont\ttfam=\tentt\def\tt{\fam\ttfam\tentt}%
  \textfont\scfam=\tencsc\def\sc{\fam\scfam\tencsc}%
  \textfont\sffam=\tensf\def\sf{\fam\sffam\tensf}%
  \textfont\mibfam=\tenmib%
  \textfont\sybfam=\tensyb%
  \bls{12pt}%
}

\def\elevenpoint{
  \def\rm{\fam0\elevenrm}%
  \textfont0=\elevenrm \scriptfont0=\eightrm \scriptscriptfont0=\fiverm%
  \textfont1=\eleveni  \scriptfont1=\eighti  \scriptscriptfont1=\fivei%
  \textfont2=\elevensy \scriptfont2=\eightsy \scriptscriptfont2=\fivesy%
  \textfont\itfam=\elevenit\def\it{\fam\itfam\elevenit}%
  \textfont\bffam=\elevenbf%
    \scriptfont\bffam=\eightbf%
      \scriptscriptfont\bffam=\fivebf%
  \def\bf{\fam\bffam\elevenbf}%
  \textfont\slfam=\elevensl\def\sl{\fam\slfam\elevensl}%
  \textfont\ttfam=\eleventt\def\tt{\fam\ttfam\eleventt}%
  \textfont\scfam=\elevencsc\def\sc{\fam\scfam\elevencsc}%
  \textfont\sffam=\elevensf\def\sf{\fam\sffam\elevensf}%
  \textfont\mibfam=\elevenmib%
  \textfont\sybfam=\elevensyb%
  \bls{13pt}%
}

\def\fourteenpoint{
  \def\rm{\fam0\fourteenrm}%
  \textfont0\fourteenrm  \scriptfont0\tenrm  \scriptscriptfont0\sevenrm%
  \textfont1\fourteeni   \scriptfont1\teni   \scriptscriptfont1\seveni%
  \textfont2\fourteensy  \scriptfont2\tensy  \scriptscriptfont2\sevensy%
  \textfont\itfam=\fourteenit\def\it{\fam\itfam\fourteenit}%
  \textfont\bffam=\fourteenbf%
    \scriptfont\bffam=\tenbf%
      \scriptscriptfont\bffam=\sevenbf%
  \def\bf{\fam\bffam\fourteenbf}%
  \textfont\slfam=\fourteensl\def\sl{\fam\slfam\fourteensl}%
  \textfont\ttfam=\fourteentt\def\tt{\fam\ttfam\fourteentt}%
  \textfont\scfam=\fourteencsc\def\sc{\fam\scfam\fourteencsc}%
  \textfont\sffam=\fourteensf\def\sf{\fam\sffam\fourteensf}%
  \textfont\mibfam=\fourteenmib%
  \textfont\sybfam=\fourteensyb%
  \bls{17pt}%
}

\def\seventeenpoint{
  \def\rm{\fam0\seventeenrm}%
  \textfont0\seventeenrm  \scriptfont0\elevenrm  \scriptscriptfont0\ninerm%
  \textfont1\seventeeni   \scriptfont1\eleveni   \scriptscriptfont1\ninei%
  \textfont2\seventeensy  \scriptfont2\elevensy  \scriptscriptfont2\ninesy%
  \textfont\itfam=\seventeenit\def\it{\fam\itfam\seventeenit}%
  \textfont\bffam=\seventeenbf%
    \scriptfont\bffam=\elevenbf%
      \scriptscriptfont\bffam=\ninebf%
  \def\bf{\fam\bffam\seventeenbf}%
  \textfont\slfam=\seventeensl\def\sl{\fam\slfam\seventeensl}%
  \textfont\ttfam=\seventeentt\def\tt{\fam\ttfam\seventeentt}%
  \textfont\scfam=\seventeencsc\def\sc{\fam\scfam\seventeencsc}%
  \textfont\sffam=\seventeensf\def\sf{\fam\sffam\seventeensf}%
  \textfont\mibfam=\seventeenmib%
  \textfont\sybfam=\seventeensyb%
  \bls{20pt}%
}

\lineskip=1pt      \normallineskip=\lineskip
\lineskiplimit=0pt \normallineskiplimit=\lineskiplimit




\def\Nulle{0}  
\def\Aue{1}    
\def\Afe{2}    
\def\Sue{4}    
\def\Hae{5}    
\def\Hbe{6}    
\def\Hce{7}    
\def\Hde{8}    
\def\Kwe{9}    
\def\Txe{10}   
\def\Lie{11}   
\def\Bbe{12}   


\newdimen\DimenA
\newbox\BoxA

\newcount\LastMac \LastMac=\Nulle
\newcount\HeaderNumber \HeaderNumber=0
\newcount\DefaultHeader \DefaultHeader=\HeaderNumber
\newskip\Indent

\newskip\half      \half=5.5pt plus 1.5pt minus 2.25pt
\newskip\one       \one=11pt plus 3pt minus 5.5pt
\newskip\onehalf   \onehalf=16.5pt plus 5.5pt minus 8.25pt
\newskip\two       \two=22pt plus 5.5pt minus 11pt

\def\Half{\vskip-\lastskip\vskip\half}
\def\One{\vskip-\lastskip\vskip\one}
\def\OneHalf{\vskip-\lastskip\vskip\onehalf}
\def\Two{\vskip-\lastskip\vskip\two}


\def\rTenPT{10pt plus \Feathering}

\def\TenPT{10pt plus \Feathering} 
\def\ElevenPT{11pt plus \Feathering}

\def\Raggedright{
 \rightskip=0pt plus \hsize
}

\def\Fullout{
\rightskip=0pt
}

\def\Hang#1#2{
 \hangindent=#1
 \hangafter=#2
}

\def\EveryMac{
 \Fullout
 \everypar{}
}



\def\title#1{
 \EveryMac
 \LastMac=\Nulle
 \global\HeaderNumber=0
 \global\DefaultHeader=1
 \vbox to 1pc{\vss}
 \seventeenpoint
 \Raggedright
 \noindent \bf #1
}

\def\author#1{
 \EveryMac
 \ifnum\LastMac=\Afe \OneHalf
  \else \Two
 \fi
 \LastMac=\Aue
 \fourteenpoint
 \Raggedright
 \noindent \rm #1\par
 \vskip 3pt\relax
}

\def\affiliation#1{
 \EveryMac
 \LastMac=\Afe
 \eightpoint\bls{\TenPT}
 \Raggedright
 \noindent \it #1\par
}

\def\abstract{%
 \EveryMac
 \Two
 \LastMac=\Sue
 \everypar{\Hang{11pc}{0}}
 \noindent\ninebf ABSTRACT\par
 \tenpoint\bls{\ElevenPT}
 \Fullout
 \noindent\rm
}

\def\keywords{
 \EveryMac
 \Half
 \LastMac=\Kwe
 \everypar{\Hang{11pc}{0}}
 \tenpoint\bls{\ElevenPT}
 \Fullout
 \noindent\hbox{\bf Key words:\ }
 \rm
}


\def\maketitle{%
  \Two%
  \EndOpening%
  \MakePage%
}


\def\pageoffset#1#2{\hoffset=#1\relax\voffset=#2\relax}


\def\Autonumber{
 \global\AutoNumbertrue  
}

\newif\ifAutoNumber \AutoNumberfalse
\newcount\Sec        
\newcount\SecSec
\newcount\SecSecSec

\Sec=0

\def\:{\let\@sptoken= } \:  
\def\:{\@xifnch} \expandafter\def\: {\futurelet\@tempc\@ifnch}

\def\@ifnextchar#1#2#3{%
  \let\@tempMACe #1%
  \def\@tempMACa{#2}%
  \def\@tempMACb{#3}%
  \futurelet \@tempMACc\@ifnch%
}

\def\@ifnch{%
\ifx \@tempMACc \@sptoken%
  \let\@tempMACd\@xifnch%
\else%
  \ifx \@tempMACc \@tempMACe%
    \let\@tempMACd\@tempMACa%
  \else%
    \let\@tempMACd\@tempMACb%
  \fi%
\fi%
\@tempMACd%
}

\def\@ifstar#1#2{\@ifnextchar *{\def\@tempMACa*{#1}\@tempMACa}{#2}}

\def\section{\@ifstar{\@ssection}{\@section}}

\def\@section#1{
 \EveryMac
 \Two
 \LastMac=\Hae
 \ninepoint\bls{\ElevenPT}
 \bf
 \Raggedright
 \ifAutoNumber
  \advance\Sec by 1
  \noindent\number\Sec\hskip 1pc \uppercase{#1}
  \SecSec=0
 \else
  \noindent \uppercase{#1}
 \fi
 \nobreak
}

\def\@ssection#1{
 \EveryMac
 \ifnum\LastMac=\Hae \Half
  \else \OneHalf
 \fi
 \LastMac=\Hae
 \tenpoint\bls{\ElevenPT}
 \bf
 \Raggedright
 \noindent\uppercase{#1}
}

\def\subsection#1{
 \EveryMac
 \ifnum\LastMac=\Hae \Half
  \else \OneHalf
 \fi
 \LastMac=\Hbe
 \tenpoint\bls{\ElevenPT}
 \bf
 \Raggedright
 \ifAutoNumber
  \advance\SecSec by 1
  \noindent\number\Sec.\number\SecSec
  \hskip 1pc #1
  \SecSecSec=0
 \else
  \noindent #1
 \fi
 \nobreak
}

\def\subsubsection#1{
 \EveryMac
 \ifnum\LastMac=\Hbe \Half
  \else \OneHalf
 \fi
 \LastMac=\Hce
 \ninepoint\bls{\ElevenPT}
 \it
 \Raggedright
 \ifAutoNumber
  \advance\SecSecSec by 1
  \noindent\number\Sec.\number\SecSec.\number\SecSecSec
  \hskip 1pc #1
 \else
  \noindent #1
 \fi
 \nobreak
}

\def\paragraph#1{
 \EveryMac
 \One
 \LastMac=\Hde
 \ninepoint\bls{\ElevenPT}
 \noindent \it #1
 \rm
}


\def\tx{
 \EveryMac
 \ifnum\LastMac=\Lie \Half\fi
 \ifnum\LastMac=\Hae \nobreak\Half\fi
 \ifnum\LastMac=\Hbe \nobreak\Half\fi
 \ifnum\LastMac=\Hce \nobreak\Half\fi
 \ifnum\LastMac=\Lie \else \noindent\fi
 \LastMac=\Txe
 \ninepoint\bls{\ElevenPT}
 \rm
}


\def\item{
 \par
 \EveryMac
 \ifnum\LastMac=\Lie
  \else \Half
 \fi
 \LastMac=\Lie
 \ninepoint\bls{\ElevenPT}
 \rm
}


\def\bibitem{
 \par
 \EveryMac
 \ifnum\LastMac=\Bbe
  \else \Half
 \fi
 \LastMac=\Bbe
 \Hang{1.5em}{1}
 \eightpoint\bls{\TenPT}
 \Raggedright
 \noindent \rm
}


\newtoks\CatchLine

\def\@journal{Mon.\ Not.\ R.\ Astron.\ Soc.\ }  
\def\@pubyear{1993}        
\def\@pagerange{000--000}  
\def\@volume{000}          
\def\@microfiche{}         %

\def\pubyear#1{\gdef\@pubyear{#1}\@makecatchline}
\def\pagerange#1{\gdef\@pagerange{#1}\@makecatchline}
\def\volume#1{\gdef\@volume{#1}\@makecatchline}
\def\microfiche#1{\gdef\@microfiche{and Microfiche\ #1}\@makecatchline}

\def\@makecatchline{%
  \global\CatchLine{%
    {\rm \@journal {\bf \@volume},\ \@pagerange\ (\@pubyear)\ \@microfiche}}%
}

\@makecatchline 

\newtoks\LeftHeader
\def\shortauthor#1{
 \global\LeftHeader{#1}
}

\newtoks\RightHeader
\def\shorttitle#1{
 \global\RightHeader{#1}
}

\def\PageHead{
 \EveryMac
 \ifnum\HeaderNumber=1 \Pagehead
  \else \Catchline
 \fi
}

\def\Catchline{%
 \vbox to 0pt{\vskip-22.5pt
  \hbox to \PageWidth{\vbox to8.5pt{}\noindent
  \eightpoint\the\CatchLine\hfill}\vss}
 \nointerlineskip
}

\def\Pagehead{%
 \ifodd\pageno
   \vbox to 0pt{\vskip-22.5pt
   \hbox to \PageWidth{\vbox to8.5pt{}\elevenpoint\it\noindent
    \hfill\the\RightHeader\hskip1.5em\rm\folio}\vss}
 \else
   \vbox to 0pt{\vskip-22.5pt
   \hbox to \PageWidth{\vbox to8.5pt{}\elevenpoint\rm\noindent
   \folio\hskip1.5em\it\the\LeftHeader\hfill}\vss}
 \fi
 \nointerlineskip
}

\def\PageFoot{} 

\def\authorcomment#1{%
  \gdef\PageFoot{%
    \nointerlineskip%
    \vbox to 22pt{\vfil%
      \hbox to \PageWidth{\elevenpoint\rm\noindent \hfil #1 \hfil}}%
  }%
}

\everydisplay{\displaysetup}

\newif\ifeqno
\newif\ifleqno

\def\displaysetup#1$${%
 \displaytest#1\eqno\eqno\displaytest
}

\def\displaytest#1\eqno#2\eqno#3\displaytest{%
 \if!#3!\ldisplaytest#1\leqno\leqno\ldisplaytest
 \else\eqnotrue\leqnofalse\def\eqn{#2}\def\eq{#1}\fi
 \generaldisplay$$}

\def\ldisplaytest#1\leqno#2\leqno#3\ldisplaytest{%
 \def\eq{#1}%
 \if!#3!\eqnofalse\else\eqnotrue\leqnotrue
  \def\eqn{#2}\fi}

\def\generaldisplay{%
\ifeqno \ifleqno 
   \hbox to \hsize{\noindent
     $\displaystyle\eq$\hfil$\displaystyle\eqn$}
  \else
    \hbox to \hsize{\noindent
     $\displaystyle\eq$\hfil$\displaystyle\eqn$}
  \fi
 \else
 \hbox to \hsize{\vbox{\noindent
  $\displaystyle\eq$\hfil}}
 \fi
}

\def\@notice{%
  \par\Two%
  \bls{12pt}%
  \noindent\tenrm This paper has been produced using the Blackwell
                  Scientific Publications \TeX\ macros.%
}

\outer\def\bye{\@notice\par\vfill\supereject\end}

\everyjob{%
  \Warn{Monthly notices of the RAS journal style (\@typeface)\space
        v\@version,\space \@verdate.}\Warn{}%
}




\newif\if@debug \@debugfalse  

\def\Print#1{\if@debug\immediate\write16{#1}\else \fi}
\def\Warn#1{\immediate\write16{#1}}
\def\wlog#1{}

\newcount\Iteration 

\newif\ifFigureBoxes  
\FigureBoxestrue

\def\Single{0} \def\Double{1}                 
\def\Figure{0} \def\Table{1}                  

\def\InStack{0}  
\def\InZoneA{1}
\def\InZoneB{2}
\def\InZoneC{3}

\newcount\TEMPCOUNT 
\newdimen\TEMPDIMEN 
\newbox\TEMPBOX     
\newbox\VOIDBOX     

\newcount\LengthOfStack 
\newcount\MaxItems      
\newcount\StackPointer
\newcount\Point         
\newcount\NextFigure    
\newcount\NextTable     
\newcount\NextItem      

\newcount\StatusStack   
\newcount\NumStack      
\newcount\TypeStack     
\newcount\SpanStack     
\newcount\BoxStack      

\newcount\ItemSTATUS    
\newcount\ItemNUMBER    
\newcount\ItemTYPE      
\newcount\ItemSPAN      
\newbox\ItemBOX         
\newdimen\ItemSIZE      

\newdimen\PageHeight    
\newdimen\TextLeading   
\newdimen\Feathering    
\newcount\LinesPerPage  
\newdimen\ColumnWidth   
\newdimen\ColumnGap     
\newdimen\PageWidth     
\newdimen\BodgeHeight   
\newcount\Leading       

\newdimen\ZoneBSize  
\newdimen\TextSize   
\newbox\ZoneABOX     
\newbox\ZoneBBOX     
\newbox\ZoneCBOX     

\newif\ifFirstSingleItem
\newif\ifFirstZoneA
\newif\ifMakePageInComplete
\newif\ifMoreFigures \MoreFiguresfalse 
\newif\ifMoreTables  \MoreTablesfalse  

\newif\ifFigInZoneB 
\newif\ifFigInZoneC 
\newif\ifTabInZoneB 
\newif\ifTabInZoneC

\newif\ifZoneAFullPage

\newbox\MidBOX    
\newbox\LeftBOX
\newbox\RightBOX
\newbox\PageBOX   

\newif\ifLeftCOL  
\LeftCOLtrue

\newdimen\ZoneBAdjust

\newcount\ItemFits
\def\Yes{1}
\def\No{2}




\MaxItems=15
\NextFigure=0        
\NextTable=1

\BodgeHeight=6pt
\TextLeading=11pt    
\Leading=11
\Feathering=0pt      
\LinesPerPage=61     
\topskip=\TextLeading
\ColumnWidth=20pc    
\ColumnGap=2pc       

\def\ItemSep{\vskip \TextLeading plus \TextLeading minus 4pt}

\FigureBoxesfalse 

\parskip=0pt
\parindent=18pt
\widowpenalty=0
\clubpenalty=10000
\tolerance=1500
\hbadness=1500
\abovedisplayskip=6pt plus 2pt minus 2pt
\belowdisplayskip=6pt plus 2pt minus 2pt
\abovedisplayshortskip=6pt plus 2pt minus 2pt
\belowdisplayshortskip=6pt plus 2pt minus 2pt

\PageHeight=\TextLeading 
\multiply\PageHeight by \LinesPerPage
\advance\PageHeight by \topskip

\PageWidth=2\ColumnWidth
\advance\PageWidth by \ColumnGap




\newcount\DUMMY \StatusStack=\allocationnumber
\newcount\DUMMY \newcount\DUMMY \newcount\DUMMY 
\newcount\DUMMY \newcount\DUMMY \newcount\DUMMY 
\newcount\DUMMY \newcount\DUMMY \newcount\DUMMY
\newcount\DUMMY \newcount\DUMMY \newcount\DUMMY 
\newcount\DUMMY \newcount\DUMMY \newcount\DUMMY

\newcount\DUMMY \NumStack=\allocationnumber
\newcount\DUMMY \newcount\DUMMY \newcount\DUMMY 
\newcount\DUMMY \newcount\DUMMY \newcount\DUMMY 
\newcount\DUMMY \newcount\DUMMY \newcount\DUMMY 
\newcount\DUMMY \newcount\DUMMY \newcount\DUMMY 
\newcount\DUMMY \newcount\DUMMY \newcount\DUMMY

\newcount\DUMMY \TypeStack=\allocationnumber
\newcount\DUMMY \newcount\DUMMY \newcount\DUMMY 
\newcount\DUMMY \newcount\DUMMY \newcount\DUMMY 
\newcount\DUMMY \newcount\DUMMY \newcount\DUMMY 
\newcount\DUMMY \newcount\DUMMY \newcount\DUMMY 
\newcount\DUMMY \newcount\DUMMY \newcount\DUMMY

\newcount\DUMMY \SpanStack=\allocationnumber
\newcount\DUMMY \newcount\DUMMY \newcount\DUMMY 
\newcount\DUMMY \newcount\DUMMY \newcount\DUMMY 
\newcount\DUMMY \newcount\DUMMY \newcount\DUMMY 
\newcount\DUMMY \newcount\DUMMY \newcount\DUMMY 
\newcount\DUMMY \newcount\DUMMY \newcount\DUMMY

\newbox\DUMMY   \BoxStack=\allocationnumber
\newbox\DUMMY   \newbox\DUMMY \newbox\DUMMY 
\newbox\DUMMY   \newbox\DUMMY \newbox\DUMMY 
\newbox\DUMMY   \newbox\DUMMY \newbox\DUMMY 
\newbox\DUMMY   \newbox\DUMMY \newbox\DUMMY 
\newbox\DUMMY   \newbox\DUMMY \newbox\DUMMY

\def\wlog{\immediate\write-1}


\def\GetItemAll#1{%
 \GetItemSTATUS{#1}
 \GetItemNUMBER{#1}
 \GetItemTYPE{#1}
 \GetItemSPAN{#1}
 \GetItemBOX{#1}
}

\def\GetItemSTATUS#1{%
 \Point=\StatusStack
 \advance\Point by #1
 \global\ItemSTATUS=\count\Point
}

\def\GetItemNUMBER#1{%
 \Point=\NumStack
 \advance\Point by #1
 \global\ItemNUMBER=\count\Point
}

\def\GetItemTYPE#1{%
 \Point=\TypeStack
 \advance\Point by #1
 \global\ItemTYPE=\count\Point
}

\def\GetItemSPAN#1{%
 \Point\SpanStack
 \advance\Point by #1
 \global\ItemSPAN=\count\Point
}

\def\GetItemBOX#1{%
 \Point=\BoxStack
 \advance\Point by #1
 \global\setbox\ItemBOX=\vbox{\copy\Point}
 \global\ItemSIZE=\ht\ItemBOX
 \global\advance\ItemSIZE by \dp\ItemBOX
 \TEMPCOUNT=\ItemSIZE
 \divide\TEMPCOUNT by \Leading
 \divide\TEMPCOUNT by 65536
 \advance\TEMPCOUNT by 1
 \ItemSIZE=\TEMPCOUNT pt
 \global\multiply\ItemSIZE by \Leading
}


\def\JoinStack{%
 \ifnum\LengthOfStack=\MaxItems 
  \Warn{WARNING: Stack is full...some items will be lost!}
 \else
  \Point=\StatusStack
  \advance\Point by \LengthOfStack
  \global\count\Point=\ItemSTATUS
  \Point=\NumStack
  \advance\Point by \LengthOfStack
  \global\count\Point=\ItemNUMBER
  \Point=\TypeStack
  \advance\Point by \LengthOfStack
  \global\count\Point=\ItemTYPE
  \Point\SpanStack
  \advance\Point by \LengthOfStack
  \global\count\Point=\ItemSPAN
  \Point=\BoxStack
  \advance\Point by \LengthOfStack
  \global\setbox\Point=\vbox{\copy\ItemBOX}
  \global\advance\LengthOfStack by 1
  \ifnum\ItemTYPE=\Figure 
   \global\MoreFigurestrue
  \else
   \global\MoreTablestrue
  \fi
 \fi
}


\def\LeaveStack#1{%
 {\Iteration=#1
 \loop
 \ifnum\Iteration<\LengthOfStack
  \advance\Iteration by 1
  \GetItemSTATUS{\Iteration}
   \advance\Point by -1
   \global\count\Point=\ItemSTATUS
  \GetItemNUMBER{\Iteration}
   \advance\Point by -1
   \global\count\Point=\ItemNUMBER
  \GetItemTYPE{\Iteration}
   \advance\Point by -1
   \global\count\Point=\ItemTYPE
  \GetItemSPAN{\Iteration}
   \advance\Point by -1
   \global\count\Point=\ItemSPAN
  \GetItemBOX{\Iteration}
   \advance\Point by -1
   \global\setbox\Point=\vbox{\copy\ItemBOX}
 \repeat}
 \global\advance\LengthOfStack by -1
}


\newif\ifStackNotClean

\def\CleanStack{%
 \StackNotCleantrue
 {\Iteration=0
  \loop
   \ifStackNotClean
    \GetItemSTATUS{\Iteration}
    \ifnum\ItemSTATUS=\InStack
     \advance\Iteration by 1
     \else
      \LeaveStack{\Iteration}
    \fi
   \ifnum\LengthOfStack<\Iteration
    \StackNotCleanfalse
   \fi
 \repeat}
}


\def\FindItem#1#2{%
 \global\StackPointer=-1 
 {\Iteration=0
  \loop
  \ifnum\Iteration<\LengthOfStack
   \GetItemSTATUS{\Iteration}
   \ifnum\ItemSTATUS=\InStack
    \GetItemTYPE{\Iteration}
    \ifnum\ItemTYPE=#1
     \GetItemNUMBER{\Iteration}
     \ifnum\ItemNUMBER=#2
      \global\StackPointer=\Iteration
      \Iteration=\LengthOfStack 
     \fi
    \fi
   \fi
  \advance\Iteration by 1
 \repeat}
}


\def\FindNext{%
 \global\StackPointer=-1 
 {\Iteration=0
  \loop
  \ifnum\Iteration<\LengthOfStack
   \GetItemSTATUS{\Iteration}
   \ifnum\ItemSTATUS=\InStack
    \GetItemTYPE{\Iteration}
   \ifnum\ItemTYPE=\Figure
    \ifMoreFigures
      \global\NextItem=\Figure
      \global\StackPointer=\Iteration
      \Iteration=\LengthOfStack 
    \fi
   \fi
   \ifnum\ItemTYPE=\Table
    \ifMoreTables
      \global\NextItem=\Table
      \global\StackPointer=\Iteration
      \Iteration=\LengthOfStack 
    \fi
   \fi
  \fi
  \advance\Iteration by 1
 \repeat}
}


\def\ChangeStatus#1#2{%
 \Point=\StatusStack
 \advance\Point by #1
 \global\count\Point=#2
}



\def\Zone{\InZoneA}

\ZoneBAdjust=0pt

\def\MakePage{
 \global\ZoneBSize=\PageHeight
 \global\TextSize=\ZoneBSize
 \global\ZoneAFullPagefalse
 \global\topskip=\TextLeading
 \MakePageInCompletetrue
 \MoreFigurestrue
 \MoreTablestrue
 \FigInZoneBfalse
 \FigInZoneCfalse
 \TabInZoneBfalse
 \TabInZoneCfalse
 \global\FirstSingleItemtrue
 \global\FirstZoneAtrue
 \global\setbox\ZoneABOX=\box\VOIDBOX
 \global\setbox\ZoneBBOX=\box\VOIDBOX
 \global\setbox\ZoneCBOX=\box\VOIDBOX
 \loop
  \ifMakePageInComplete
 \FindNext
 \ifnum\StackPointer=-1
  \NextItem=-1
  \MoreFiguresfalse
  \MoreTablesfalse
 \fi
 \ifnum\NextItem=\Figure
   \FindItem{\Figure}{\NextFigure}
   \ifnum\StackPointer=-1 \global\MoreFiguresfalse
   \else
    \GetItemSPAN{\StackPointer}
    \ifnum\ItemSPAN=\Single \def\Zone{\InZoneB}\relax
     \ifFigInZoneC \global\MoreFiguresfalse\fi
    \else
     \def\Zone{\InZoneA}
     \ifFigInZoneB \def\Zone{\InZoneC}\fi
    \fi
   \fi
   \ifMoreFigures\Print{}\FigureItems\fi
 \fi
\ifnum\NextItem=\Table
   \FindItem{\Table}{\NextTable}
   \ifnum\StackPointer=-1 \global\MoreTablesfalse
   \else
    \GetItemSPAN{\StackPointer}
    \ifnum\ItemSPAN=\Single\relax
     \ifTabInZoneC \global\MoreTablesfalse\fi
    \else
     \def\Zone{\InZoneA}
     \ifTabInZoneB \def\Zone{\InZoneC}\fi
    \fi
   \fi
   \ifMoreTables\Print{}\TableItems\fi
 \fi
   \MakePageInCompletefalse 
   \ifMoreFigures\MakePageInCompletetrue\fi
   \ifMoreTables\MakePageInCompletetrue\fi
 \repeat
 \ifZoneAFullPage
  \global\TextSize=0pt
  \global\ZoneBSize=0pt
  \global\vsize=0pt\relax
  \global\topskip=0pt\relax
  \vbox to 0pt{\vss}
  \eject
 \else
 \global\advance\ZoneBSize by -\ZoneBAdjust
 \global\vsize=\ZoneBSize
 \global\hsize=\ColumnWidth
 \global\ZoneBAdjust=0pt
 \ifdim\TextSize<23pt
 \Warn{}
 \Warn{* Making column fall short: TextSize=\the\TextSize *}
 \vskip-\lastskip\eject\fi
 \fi
}

\def\MakeRightCol{
 \global\TextSize=\ZoneBSize
 \MakePageInCompletetrue
 \MoreFigurestrue
 \MoreTablestrue
 \global\FirstSingleItemtrue
 \global\setbox\ZoneBBOX=\box\VOIDBOX
 \def\Zone{\InZoneB}
 \loop
  \ifMakePageInComplete
 \FindNext
 \ifnum\StackPointer=-1
  \NextItem=-1
  \MoreFiguresfalse
  \MoreTablesfalse
 \fi
 \ifnum\NextItem=\Figure
   \FindItem{\Figure}{\NextFigure}
   \ifnum\StackPointer=-1 \MoreFiguresfalse
   \else
    \GetItemSPAN{\StackPointer}
    \ifnum\ItemSPAN=\Double\relax
     \MoreFiguresfalse\fi
   \fi
   \ifMoreFigures\Print{}\FigureItems\fi
 \fi
 \ifnum\NextItem=\Table
   \FindItem{\Table}{\NextTable}
   \ifnum\StackPointer=-1 \MoreTablesfalse
   \else
    \GetItemSPAN{\StackPointer}
    \ifnum\ItemSPAN=\Double\relax
     \MoreTablesfalse\fi
   \fi
   \ifMoreTables\Print{}\TableItems\fi
 \fi
   \MakePageInCompletefalse 
   \ifMoreFigures\MakePageInCompletetrue\fi
   \ifMoreTables\MakePageInCompletetrue\fi
 \repeat
 \ifZoneAFullPage
  \global\TextSize=0pt
  \global\ZoneBSize=0pt
  \global\vsize=0pt\relax
  \global\topskip=0pt\relax
  \vbox to 0pt{\vss}
  \eject
 \else
 \global\vsize=\ZoneBSize
 \global\hsize=\ColumnWidth
 \ifdim\TextSize<23pt
 \Warn{}
 \Warn{* Making column fall short: TextSize=\the\TextSize *}
 \vskip-\lastskip\eject\fi
\fi
}

\def\FigureItems{
 \Print{Considering...}
 \ShowItem{\StackPointer}
 \GetItemBOX{\StackPointer} 
 \GetItemSPAN{\StackPointer}
  \CheckFitInZone 
  \ifnum\ItemFits=\Yes
   \ifnum\ItemSPAN=\Single
     \ChangeStatus{\StackPointer}{\InZoneB} 
     \global\FigInZoneBtrue
     \ifFirstSingleItem
      \hbox{}\vskip-\BodgeHeight
     \global\advance\ItemSIZE by \TextLeading
     \fi
     \unvbox\ItemBOX\ItemSep
     \global\FirstSingleItemfalse
     \global\advance\TextSize by -\ItemSIZE
     \global\advance\TextSize by -\TextLeading
   \else
    \ifFirstZoneA
     \global\advance\ItemSIZE by \TextLeading
     \global\FirstZoneAfalse\fi
    \global\advance\TextSize by -\ItemSIZE
    \global\advance\TextSize by -\TextLeading
    \global\advance\ZoneBSize by -\ItemSIZE
    \global\advance\ZoneBSize by -\TextLeading
    \ifFigInZoneB\relax
     \else
     \ifdim\TextSize<3\TextLeading
     \global\ZoneAFullPagetrue
     \fi
    \fi
    \ChangeStatus{\StackPointer}{\Zone}
    \ifnum\Zone=\InZoneC \global\FigInZoneCtrue\fi
  \fi
   \Print{TextSize=\the\TextSize}
   \Print{ZoneBSize=\the\ZoneBSize}
  \global\advance\NextFigure by 1
   \Print{This figure has been placed.}
  \else
   \Print{No space available for this figure...holding over.}
   \Print{}
   \global\MoreFiguresfalse
  \fi
}

\def\TableItems{
 \Print{Considering...}
 \ShowItem{\StackPointer}
 \GetItemBOX{\StackPointer} 
 \GetItemSPAN{\StackPointer}
  \CheckFitInZone 
  \ifnum\ItemFits=\Yes
   \ifnum\ItemSPAN=\Single
    \ChangeStatus{\StackPointer}{\InZoneB}
     \global\TabInZoneBtrue
     \ifFirstSingleItem
      \hbox{}\vskip-\BodgeHeight
     \global\advance\ItemSIZE by \TextLeading
     \fi
     \unvbox\ItemBOX\ItemSep
     \global\FirstSingleItemfalse
     \global\advance\TextSize by -\ItemSIZE
     \global\advance\TextSize by -\TextLeading
   \else
    \ifFirstZoneA
    \global\advance\ItemSIZE by \TextLeading
    \global\FirstZoneAfalse\fi
    \global\advance\TextSize by -\ItemSIZE
    \global\advance\TextSize by -\TextLeading
    \global\advance\ZoneBSize by -\ItemSIZE
    \global\advance\ZoneBSize by -\TextLeading
    \ifFigInZoneB\relax
     \else
     \ifdim\TextSize<3\TextLeading
     \global\ZoneAFullPagetrue
     \fi
    \fi
    \ChangeStatus{\StackPointer}{\Zone}
    \ifnum\Zone=\InZoneC \global\TabInZoneCtrue\fi
   \fi
  \global\advance\NextTable by 1
   \Print{This table has been placed.}
  \else
  \Print{No space available for this table...holding over.}
   \Print{}
   \global\MoreTablesfalse
  \fi
}


\def\CheckFitInZone{%
{\advance\TextSize by -\ItemSIZE
 \advance\TextSize by -\TextLeading
 \ifFirstSingleItem
  \advance\TextSize by \TextLeading
 \fi
 \ifnum\Zone=\InZoneA\relax
  \else \advance\TextSize by -\ZoneBAdjust
 \fi
 \ifdim\TextSize<3\TextLeading \global\ItemFits=\No
 \else \global\ItemFits=\Yes\fi}
}

\def\BF#1#2{
 \ItemSTATUS=\InStack
 \ItemNUMBER=#1
 \ItemTYPE=\Figure
 \if#2S \ItemSPAN=\Single
  \else \ItemSPAN=\Double
 \fi
 \setbox\ItemBOX=\vbox{}
}

\def\BT#1#2{
 \ItemSTATUS=\InStack
 \ItemNUMBER=#1
 \ItemTYPE=\Table
 \if#2S \ItemSPAN=\Single
  \else \ItemSPAN=\Double
 \fi
 \setbox\ItemBOX=\vbox{}
}

\def\BeginOpening{%
 \hsize=\PageWidth
 \global\setbox\ItemBOX=\vbox\bgroup
}

\let\begintopmatter=\BeginOpening  

\def\EndOpening{%
 \egroup
 \ItemNUMBER=0
 \ItemTYPE=\Figure
 \ItemSPAN=\Double
 \ItemSTATUS=\InStack
 \JoinStack
}


\newbox\tmpbox

\def\FC#1#2#3#4{%
  \ItemSTATUS=\InStack
  \ItemNUMBER=#1
  \ItemTYPE=\Figure
  \if#2S
    \ItemSPAN=\Single \TEMPDIMEN=\ColumnWidth
  \else
    \ItemSPAN=\Double \TEMPDIMEN=\PageWidth
  \fi
  {\hsize=\TEMPDIMEN
   \global\setbox\ItemBOX=\vbox{%
     \ifFigureBoxes
       \B{\TEMPDIMEN}{#3}
     \else
       \vbox to #3{\vfil}%
     \fi%
     \eightpoint\rm\bls{\rTenPT}%
     \vskip 5.5pt plus 6pt%
     \setbox\tmpbox=\vbox{#4\par}%
     \ifdim\ht\tmpbox>10pt 
       \noindent #4\par%
     \else
       \hbox to \hsize{\hfil #4\hfil}%
     \fi%
   }%
  }%
  \JoinStack%
  \Print{Processing source for figure {\the\ItemNUMBER}}%
}

\let\figure=\FC  

\def\TH#1#2#3#4{%
 \ItemSTATUS=\InStack
 \ItemNUMBER=#1
 \ItemTYPE=\Table
 \if#2S \ItemSPAN=\Single \TEMPDIMEN=\ColumnWidth
  \else \ItemSPAN=\Double \TEMPDIMEN=\PageWidth
 \fi
{\hsize=\TEMPDIMEN
\eightpoint\bls{\rTenPT}\rm
\global\setbox\ItemBOX=\vbox{\noindent#3\vskip 5.5pt plus5.5pt\noindent#4}}
 \JoinStack
 \Print{Processing source for table {\the\ItemNUMBER}}
}


\def\UnloadZoneA{%
\FirstZoneAtrue
 \Iteration=0
  \loop
   \ifnum\Iteration<\LengthOfStack
    \GetItemSTATUS{\Iteration}
    \ifnum\ItemSTATUS=\InZoneA
     \GetItemBOX{\Iteration}
     \ifFirstZoneA \vbox to \BodgeHeight{\vfil}%
     \FirstZoneAfalse\fi
     \unvbox\ItemBOX\ItemSep
     \LeaveStack{\Iteration}
     \else
     \advance\Iteration by 1
   \fi
 \repeat
}

\def\UnloadZoneC{%
\Iteration=0
  \loop
   \ifnum\Iteration<\LengthOfStack
    \GetItemSTATUS{\Iteration}
    \ifnum\ItemSTATUS=\InZoneC
     \GetItemBOX{\Iteration}
     \ItemSep\unvbox\ItemBOX
     \LeaveStack{\Iteration}
     \else
     \advance\Iteration by 1
   \fi
 \repeat
}


\def\ShowItem#1{
  {\GetItemAll{#1}
  \Print{\the#1:
  {TYPE=\ifnum\ItemTYPE=\Figure Figure\else Table\fi}
  {NUMBER=\the\ItemNUMBER}
  {SPAN=\ifnum\ItemSPAN=\Single Single\else Double\fi}
  {SIZE=\the\ItemSIZE}}}
}

\def\ShowStack{%
 \Print{}
 \Print{LengthOfStack = \the\LengthOfStack}
 \ifnum\LengthOfStack=0 \Print{Stack is empty}\fi
 \Iteration=0
 \loop
 \ifnum\Iteration<\LengthOfStack
  \ShowItem{\Iteration}
  \advance\Iteration by 1
 \repeat
}

\def\B#1#2{%
\hbox{\vrule\kern-0.4pt\vbox to #2{%
\hrule width #1\vfill\hrule}\kern-0.4pt\vrule}
}

\def\Ref#1{\begingroup\global\setbox\TEMPBOX=\vbox{\hsize=2in\noindent#1}\endgroup
\ht1=0pt\dp1=0pt\wd1=0pt\vadjust{\vtop to 0pt{\advance
\hsize0.5pc\kern-10pt\moveright\hsize\box\TEMPBOX\vss}}}

\def\MarkRef#1{\leavevmode\thinspace\hbox{\vrule\vtop
{\vbox{\hrule\kern1pt\hbox{\vphantom{\rm/}\thinspace{\rm#1}%
\thinspace}}\kern1pt\hrule}\vrule}\thinspace}%


\output{%
 \ifLeftCOL
  \global\setbox\LeftBOX=\vbox to \ZoneBSize{\box255\unvbox\ZoneBBOX}
  \global\LeftCOLfalse
  \MakeRightCol
 \else
  \setbox\RightBOX=\vbox to \ZoneBSize{\box255\unvbox\ZoneBBOX}
  \setbox\MidBOX=\hbox{\box\LeftBOX\hskip\ColumnGap\box\RightBOX}
  \setbox\PageBOX=\vbox to \PageHeight{%
  \UnloadZoneA\box\MidBOX\UnloadZoneC}
  \shipout\vbox{\PageHead\box\PageBOX\PageFoot}
  \global\advance\pageno by 1
  \global\HeaderNumber=\DefaultHeader
  \global\LeftCOLtrue
  \CleanStack
  \MakePage
 \fi
}


\catcode `\@=12 


\input psfig


\def\simless{\mathbin{\lower 3pt\hbox
   {$\rlap{\raise 5pt\hbox{$\char'074$}}\mathchar"7218$}}}   
\def\simgreat{\mathbin{\lower 3pt\hbox
   {$\rlap{\raise 5pt\hbox{$\char'076$}}\mathchar"7218$}}}   
\def \AA    #1 {A\&A, #1, }
\def \AJ    #1 {AJ, #1, }
\def \ApJ   #1 {ApJ, #1, }
\def \MNRAS #1 {MNRAS, #1, }
\def \Nat   #1 {Nature, #1, }

\def \Epot { E_{\rm pot}}

\def \Ekin {E_{\rm kin}}

\def \Rg {R_{\rm g}}

\def\Rcl {R_{\rm cl}}
\def\Rj {R_{\rm J}}

\def\Mj {M_{\rm J}}
\def\tff {t_{\rm ff}}

\def\etal{{\rm et al.}}

\def\solmas{{M$_\odot$}}
\def\solm{{M_\odot}}
\def\solrad{{R$_\odot$}}

\def\Rc {R_{\rm core}}
\def\Ms {M_{\rm stars}}
\def\Rcl {R$_{\rm core}$}
\def\Mst {M$_{\rm stars}$}
\def\Mc {M_{\rm core}}
\def\Mcr {M$_{\rm core}$}
\def\Mg {M_{\rm gas}}
\def\Mgs {M$_{\rm gas}$}

\def\tcr {t$_{\rm cross}$}
\def\tc {t_{\rm cross}}
\def\tacc {t$_{\rm acc}$}
\def\tac {t_{\rm acc}}
\def\tcoll {t$_{\rm coll}$}
\def\tcol {t_{\rm coll}}

\def\vdisp {v$_{\rm disp}$}
\def\vdsp {v_{\rm disp}}
\def\Etot {E_{\rm tot}}
\def\Evir {E_{\rm vir}}
\def\rm {R_{\rm merge}}
\def\rs {R_{\star}}
\def\ms {M_{\star}}
\def\msdot {\dot{M}_{\star}}

\def\mcdot {\dot{M}_{\rm clust}}
\def\thetaonec {\theta^{1}{\rm C}}

\pageoffset{-2.5pc}{0pc}
\Autonumber


\pubyear{1998}

\begintopmatter
%

\title{On the formation of massive stars}
\author{Ian A. Bonnell$^1$, Matthew R. Bate$^2$ and Hans Zinnecker$^3$}

\affiliation{$^1$ Institute of Astronomy, Madingley Road,
Cambridge CB3 0HA}
\affiliation{$^2$ Max-Planck-Institut f\"ur Astronomie, K\"onigstuhl 17,
D-69117 Heidelberg, Germany}
\affiliation{$^3$ Astrophysikalisches Institut Potsdam, An der Sternwarte 16, D-14482 Potsdam, Germany}

\shortauthor{I. A. Bonnell et. al.}
\shorttitle{Formation of massive stars}


\abstract
We present a model for the formation of massive ($M \simgreat 10 \solm$) stars
through accretion-induced collisions in the cores of embedded dense stellar
clusters.  This model circumvents the problem of accreting onto a star
whose luminosity is sufficient to reverse the infall of gas. Instead,
the central core of the cluster accretes from the surrounding gas,
thereby decreasing  its radius until collisions between
individual components become significant. These components are, in
general, intermediate-mass stars that have formed through accretion
onto low-mass protostars. Once a sufficiently massive star has formed
to expel the remaining gas, the cluster expands in accordance with
this loss of mass, halting further collisions.  This process implies a
critical stellar density for the formation of massive stars, and
a high rate of binaries formed by tidal capture.

\keywords stars: formation -- stars: luminosity function, mass function 
-- binaries: general -- open clusters and associations: general.

\maketitle 

\section{Introduction} 

\tx Star formation on galactic and extragalactic scales is largely
concerned with the formation of massive stars, due to their radiative,
kinetic, and chemical feedback into the interstellar medium.  However,
the formation of massive stars is an, as yet, unsolved problem in
astrophysics. Although low and intermediate mass stars ($M \simless 10
\solm$) are readily explained through gravitational collapse and
subsequent accretion (Palla \& Stahler~1993), this model fails for
high-mass stars.  Radiation pressure, from a $\approx 10 \solm$ stellar
core, on the dust in the infalling gas, halts accretion
and thus limits the mass (Yorke \& Krugel~1977; Wolfire \&
Cassinelli~1987; Beech \& Mitalas~1994). The aim of this paper is to
explore an alternate model that circumvents this problem.

A valuable clue to the formation mechanism of massive stars is found
in their environments. Massive stars are rarely found in
isolation. Rather, they are predominantly found in the central regions
of rich stellar clusters (Zinnecker \etal~1993; Hillenbrand~1997).
Furthermore, even the few runaway OB stars are best explained as
having originated in the centres of dense systems (Clarke \&
Pringle~1992).

Accretion in stellar clusters is likely to be an important mechanism in
shaping the
initial mass function (Zinnecker~1982; Larson~1992).  Recent work on
the origin of the stellar mass distribution has shown that competitive
accretion in clusters naturally results in the formation of the
highest mass stars in the deepest parts of the cluster potential, ie,
the centre (Bonnell \etal~1997). The alternative, dynamical mass
segregation, is unable to account for the location of the massive
stars in such young clusters as the ONC (Bonnell \& Davies~1997;
Hillenbrand \& Hartmann~1997).  Although this model of competitive
accretion cannot apply to very massive stars, where radiation pressure
impedes the growth of stellar mass beyond 10 \solmas, it can
qualitatively account for the observed trend of the more massive stars
to be more centrally located.  In the present paper, we combine the
model of accretion onto low-mass stars in a protocluster with the
effects that such accretion has on the stellar dynamics of the
cluster.  We show that in this case stellar/ protostellar collisions
become important in the densest part of the cluster which can lead to
the build-up of very massive stars.

In Section 2, we outline why forming massive stars is problematic.
In Section 3, we discuss how the core of a cluster evolves during
accretion. Section~4 investigates when collisions become important and
Section 5 presents the models for the formation of massive stars. The
question of binaries among the massive stars is addressed in Section
6.  Section 7 explores the implications of this process while our
conclusions are presented in Section 8.

\section{The Problem} 

\tx There are a number of reasons why forming massive stars
is problematic within our current understanding of star formation.
Firstly, in isolated formation scenarios, there is the problem due to
the difficulties of accreting onto massive protostars.  The spherical
symmetric collapse of massive molecular clouds can be halted once the
protostellar core has attained masses $M \simgreat 10 \solm$. This
occurs due to the large luminosities of these stars, such that the
radiation pressure on dust, which transfers its momentum to the gas,
can halt the collapse and reverse the infall (Yorke \& Krugel~1977;
Wolfire \& Cassinelli~1987; Yorke~1993).  This limits the amount of
mass that can be directly accumulated onto a massive star. 
For accretion to continue onto a massive protostellar core, the
dust properties (abundances and sizes) have to be significantly reduced
from those found in the interstellar medium (Wolfire \& Cassinelli~1987).

This limit applies to spherically symmetric collapse and accretion.
Generally there is some angular momentum in the parent cloud and this
results in the formation of a protostellar disc (eg, Yorke,
Bodenheimer \& Laughlin~1995). Accretion from this disc onto the
massive protostar may be able to proceed if the disc is sufficiently
thin, if not the disc will be destroyed (Yorke \& Welz~1996).
Unfortunately, it is presently unclear what disc thickness is
necessary or appropriate. Still, to be able to form a massive star
this way requires that all the gas collapses to form a disc before the
protostar has gained a large fraction of its eventual mass.  The
problem with this is that collapse is non-homologous, with the
central, low angular momentum regions collapsing faster. 
Plus, gravitational torques will ensure that the disc mass
is always less than that of the central object, as they act on a timescale
of order the disc's dynamical time, much shorter than the cloud's
free-fall time, $\tff$ (Bonnell ~1994; Laughlin \& Bodenheimer~1994).
Thus, a central massive protostar should form while the
majority of the cloud is still infalling, and this limits how much
mass gathers in the disc and hence the efficiency of the process.

Even if direct collapse could form massive stars, it is difficult to
see how this process can account for the massive stars in the centre
of the dense clusters such as the Trapezium in the Orion Nebula
Cluster [ONC]. 
This is because the Jeans mass in the Trapezium protocluster cloud
must have been small, perhaps as small as 0.3 \solmas (Zinnecker \etal~1993). 
To see this, we note the Jeans length (fragmentation scale),
$$\Rj = \left({{5\Rg T}\over{2 G \mu}}\right)^{1/2} \left({4 \over 3} \pi \rho
\right)^{-1/2},\eqno(1)$$ 
must have been smaller then the average half separation of the stars
that are now present in the cluster center ("touching protostars"),
where $T$ is the temperature, $\rho$ is the density, $\Rg$ is the gas
constant, $G$ is the gravitational constant, and $\mu$ is the mean
molecular weight. From the current stellar number density ($\approx 2 - 4
\times 10^4$ stars pc$^{-3}$ [McCaughrean \& Stauffer~1994; 
Hillenbrand~1997]), we estimate $\Rj \simless 3000$ au.

We can also estimate the mass density ($\approx 100 \solm$ now in
stars) inside the current core radius ($\Rc \approx 0.05$ pc,
McCaughrean \& Stauffer~1994) of the Trapezium, yielding $\approx 2
\times 10^5$ \solmas\ pc$^{-3}$, equivalent to an initial smeared
out gas density of $\rho \approx 10^{-17} {\rm g\ cm}^{-3}$ (or more,
if the star formation efficiency, SFE, was less than 100 \% and some
gas was lost).  With $\Rj$ and $\rho$ as above, it follows then that
the Jeans mass is small, $\Mj \simless 3 \solm$ (assuming 100\% SFE),
where
$$\Mj = \left({{5R_gT}\over{2 G \mu}}\right)^{3/2} \left({4 \over 3} \pi \rho 
\right)^{-1/2}.\eqno(2)$$
This implies a temperature of 90 K.  A more realistic temperature of 20
K yields $\Mj \simless 0.3 \solm$.  To raise the initial Jeans mass to
50 \solmas, the mass of Ori $\thetaonec$, assuming the mass density of
the Trapezium cluster core, requires a  Jeans length of 8000 au or 0.04 pc
(larger than the present day star-star separations)  and
a temperature of $T \simgreat 500$ K. These values are clearly
implausibly high.

Thus, the stars that form in the centre of dense clusters should 
have initial masses much smaller than the median mass in the 
final cluster. Their initial masses are unlikely
to be larger than the minimum mass in the cluster.
In the ONC, this corresponds to an initial Jeans mass of $\Mj \approx
0.1 \solm$. Furthermore, we know that the massive stars in the centres
of young clusters had to form in situ as the clusters are generally not
old enough to allow for dynamical mass segregation (Bonnell \& Davies~1998).
To obtain the observed high masses, subsequent processes
including accretion onto the protostars, and, once $M \simgreat 10
\solm$, collisions between the intermediate mass protostars, are
required.

The problem of forming a massive star through direct collapse and
accretion also applies to scenarios where significant collisional
buildup of gaseous clumps occurs before the collapse (eg Bastien~1981;
Murray \& Lin~1996). Instead, collisions need to occur after the
objects have collapsed and attained stellar densities.  Random
encounters in star forming regions may lead to some protostellar
collisions (eg Price \& Podsiadlowski~1995) but will in general be
very rare. In order for collisions to be significant, a high density
environment is required.

\section{Accretion and cluster dynamics } 

\tx The presence of a large amount of gas in a young stellar cluster
(e.g. Lada, Strom \& Myers~1993) significantly affects the cluster's evolution.
This gas, typically comprising 1.5 to 10 times the mass in stars when
the cluster is first detectable in the IR, is a remnant of the star
formation process, and must have been significantly higher earlier on
(ie 100 \% gas initially). Accretion of this gas by individual stars
increases their masses, and, results in a spectrum of stellar masses
even from an initially uniform distribution (Bonnell \etal~1997).  

Of direct significance here is the effects accretion has on the
cluster dynamics. In the absence of large scale motions such as
rotation, the gas velocities will be uncorrelated with those of the
stars (especially in the cluster core). 
Thus, the accreted gas will, in general, have zero net momentum
compared to the stars.  Now, as each star accretes from the zero
momentum gas, it conserves momentum and thus decelerates. This
decreases the velocity dispersion in the cluster and hence decreases
its kinetic energy.  Additionally, the increase in the stellar mass
decreases the cluster's potential energy. Thus, the combined
effect of accretion in a cluster is to decrease the total energy, and
thus make it more bound. This causes the cluster to shrink to
compensate for its new, lower, total energy.

For the purposes of this paper, we consider the effects of accretion
on the dynamics of the core of a young stellar cluster.  The core is
modelled as a collection of $N$ stars within a core radius \Rcl, total
stellar mass \Mst, and total core mass \Mcr, $\Mc = \Ms + \Mg$, where
\Mgs\ is the mass in gas.  The individual components of the core are
not modelled. Instead, the core's evolution is followed by considering
its global properties and how they evolve under accretion.  
We consider the core as the central region
with a shallow radial density distribution, embedded within a larger,
steeper (such as an $R^{-2}$) distribution.  Thus, we assume
we can neglect the contribution to the potential and kinetic energies
from the surrounding cluster. We also assume that the number of stars
in the core is fixed (no stars enter or leave) and that the gas can
only enter (and not leave) the core until a massive star is formed.

In the
absence of accretion, the core will be in virial equilibrium such that
the core radius is
$$\Rc = {G\Mc\over\vdsp^2},\eqno(3)$$ where \vdisp\ is the velocity
dispersion in the core. The
primary effect of accretion is to perturb the core from this equilibrium
by decreasing the total energy. The core then adapts by shrinking,
until the velocity dispersion increases sufficiently to reestablish
virial equilibrium with a new \Rcl\ and \vdisp. These quantaties
can be calculated by first considering how the total energy decreases
as the mass is added and secondly by determining what core
radius \Rcl\ (and \vdisp) are appropriate for a virialised system
at the new, lower, total energy.

In order to calculate how the cluster core shrinks with accretion, we
consider the total energy of the stars.  The kinetic energy of the
stellar component of total mass \Mst, is $\Ekin = 1/2 \Ms \vdsp^2$,
and the potential energy of the stars from both the stellar and
gaseous components is $\Epot = -G \Ms (\Ms +
\Mg) / \Rc $.  To see the results of accretion more directly, we
express the stars' kinetic energy in terms of a characteristic stellar
momentum, $p$, defined from the velocity dispersion as, $p = \Ms
\vdsp$.  The total energy of the stars, in a cluster core of total mass \Mcr, is then given by:
$$\Etot = {p^2\over 2 \Ms} - {G \Mc \Ms \over \Rc}. \eqno(4)$$ 
From equation~4, we see that
adding mass to the core while conserving momentum decreases the total
energy.  The stars become more bound ($-G \Mc \Ms / \Rc$) as they have
a larger mass. Similarly, their kinetic energy (and velocity
dispersion) decreases due to the conservation of momentum, $p$,
($\vdsp \propto \Ms^{-1}$, $\Ekin = p^2/ 2 \Ms \propto
\Ms^{-1}$). This new lower total energy, and lower kinetic energy means
that the core is now out of virial equilibrium. It then shrinks while
conserving energy (increasing the kinetic energy and the
\vdisp) until virial equilibium is restored.  Once the cluster core
revirialises with total energy $\Evir$ (the decreased $\Etot$ after
accretion), its radius is given by
$$\Rc = {-G \Mc \Ms\over 2 \Evir},\eqno(5)$$ and the new velocity
dispersion is $$\vdsp = \sqrt{G\Mc\over\Rc}.\eqno(6)$$ 

To follow the evolution of the cluster we use equation~4 to calculate
the destabilisation of the cluster core due to the accretion, and then
equation~5 to calculate the new core radius once it has revirialised.
The two remaining unknowns that need to be considered are the timescale
on which this happens and the mass of the gas contained in the cluster core.
The minimum timescale for any process considering the dynamics of a
group of stars is the crossing time, $\tc = R/\vdsp$, where $R$ is 
a characteristic radius of the stellar distribution.

The mass of gas ($\Mg = \Mc - \Ms$) contained within the core is
uncertain and depends on the cluster dynamics and how the gas reacts
to the deepening potential. There are three main possibilities. The
first is that the accreted gas comes primarily from outside the
cluster core, such that the gas mass-fraction in the core is
constant. The second is that the accreted gas is completely contained
within the core and thus the core mass is constant while the gas mass
decreases during accretion. The last possibility is that the gas
infall onto the core is greater than the accretion onto the stars and
thus the fraction of the core mass in gas increases.  Although it is
unclear which of these scenarios is correct, we can make a qualitative
estimate based the fact that the gas infall time should be comparable
to the dynamical, or crossing, time ($\tc = R/\vdsp =
\sqrt{R^3/GM}$).  In the near-uniform density core of the cluster, 
$M \propto R^3$, the crossing time does not change drastically with
radius so the gas has time to catch up so the gas can fall in at the
same rate as the core's radius decreases.  In the surrounding areas
with a steeper ($R^{-2}$) density profile, $M \propto R$, the crossing
time increases rapidly with increasing $R$, and gas that gets left
behind cannot catch up to the core. In reality, the core can only
shrink with accretion so it will wait for the gas from outlying areas
to fall into it, but it does mean that the core is never overwhelmed
by gas from outside, it will adapt as quickly as the gas is fed in. We
assume that no gas can leave the core; that no dominant outflows
capable of removing a significant fraction of the gas occur before the
massive stars are formed. Thus, any gas which is initially in the core
will remain in the core until accreted. Furthermore, as the
accretion-induced contraction of the cluster requires several initial
crossing times (see below), significant amounts of gas will fall in
from outside the core. Therefore, for the rest of this paper, we
generally assume that the gas primarily comes from outside the core,
and that the core gas mass is proportional to the stellar mass. We do,
however, comment on how the different scenarios affect the results.

The rate at which the cluster core shrinks with added mass depends on
how the accretion time, \tacc, compares to the crossing time, \tcr, as
the core will revirialise on the order of its crossing time. For
example, if the mass is added on timescales $\tac << \tc$, no matter
how much mass is added, the cluster core will revirialise at no less
than half its original radius.  

\subsection{Slow Accretion}

\tx For slow mass accretion, $\tac >>
\tc$, the core adapts practically instantaneously. In this case, the change
in radius as a function of accreted mass, can be derived by
differentiating both equations 4 and 5 with respect to mass,
remembering that the change in $\Etot$ with mass occurs at a constant
radius whereas the change in $\Evir$ reflects changes in both mass and
radius, and that momentum is conserved during accretion. When the core
revirialises, the change in $\Evir$ and $\Etot$ are equal:
$$ \biggl({\partial \Etot \over \partial \Ms}\biggr)_{\Rc} = \biggl({d \Evir \over d \Ms}\biggr).\eqno(7)$$
Using the fact that $p^2/\Ms = G\Mc\Ms/\Rc$ when virialised, 
and assuming $\Mc \propto \Ms$,
the core radius can be shown to decrease due to accretion as:
$$\Rc \propto \Ms^{-3}.\eqno(8)$$

Alternatively, if the core mass is assumed to be constant and the
accreted gas is initially fully contained within the cluster core,
then $\Mc = \Ms + \Mg$ is constant and the cluster shrinks 
due to the loss of specific kinetic energy. In this case  $$\Rc
\propto \Ms^{-2}.\eqno(9)$$ In the third scenario where the infall of gas is 
such that it more than compensates for the accretion, then $\Mc
\propto \Mg$, and the cluster core shrinks primarily due to the
decrease in potential energy. In this case $$\Rc \propto \Mc^{-1}.\eqno(10)$$
The decrease in \Rcl\ in this case will then be from a combination of the
gas infall rate (equation~10) and the stellar mass accretion rate (equation~8).

In all cases, it is clear that even small amounts of accreted mass can
significantly decrease the core radius, due to the decrease in total
energy (equation~4), and this increases the stellar density. This is important
because it dramatically increases the rate of collisions (see \S 4
below).

\subsection{Fast Accretion}

\vbox{
\figure{1}{S}{0mm}{\vskip-0.25truein\centerline{\vbox{\psfig{figure=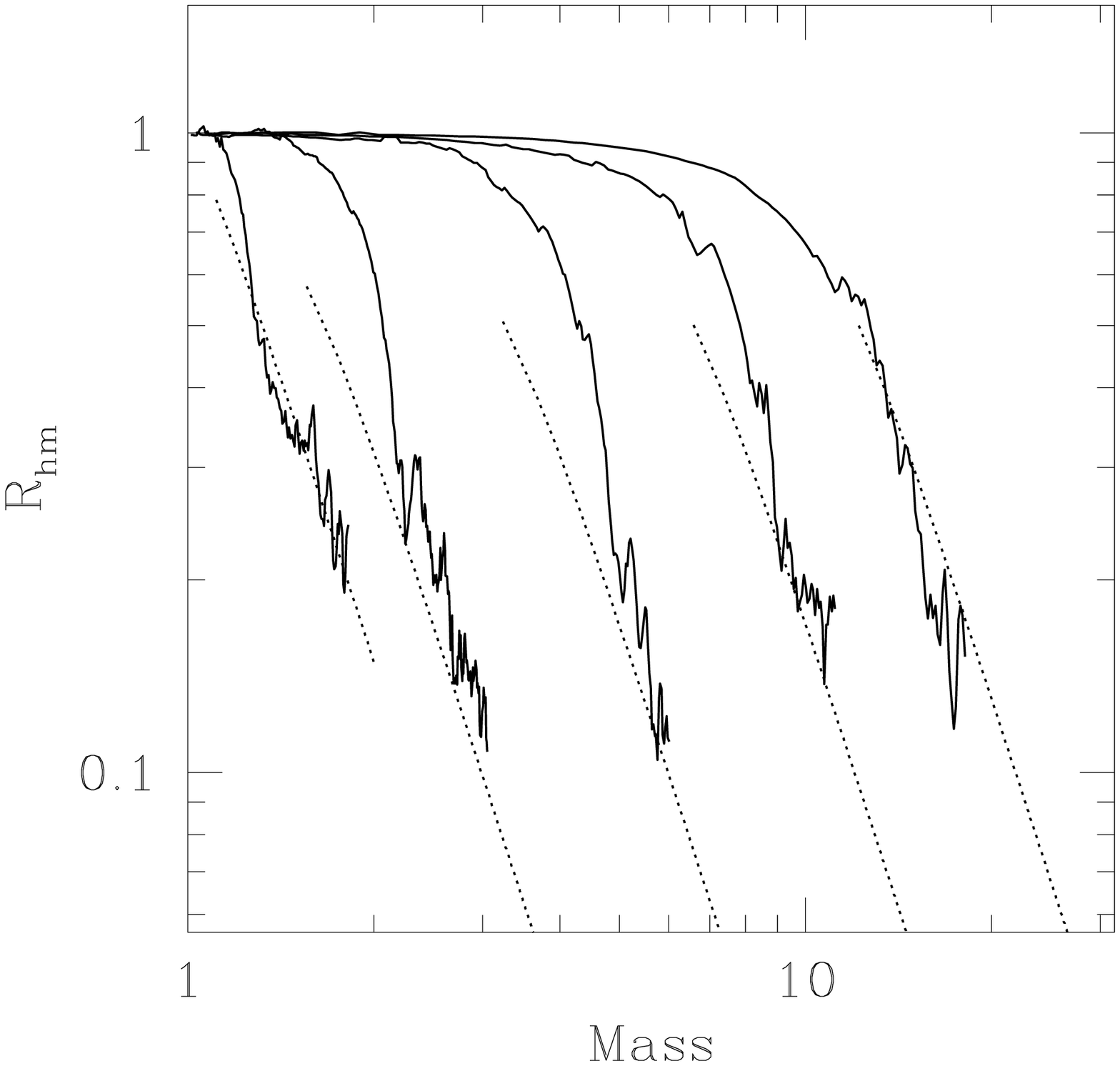,width=3.truein,height=3.truein,rwidth=3.truein,rheight=3.truein}}} 
\break\noindent
{\bf Figure 1.} The half-mass radius, in units of the initial
half-mass radius, versus total stellar mass, in units of the initial
cluster mass, for the nbody simulations (solid lines) and the toy
cluster model (dotted lines). The lines are for (from left to right)
mean stellar accretion rates of $1 \times 10^{-7}$, $5 \times
10^{-7}$, $2 \times 10^{-6}$, $5
\times 10^{-6}$ and $1 \times 10^{-5}$ \solmas yr$^{-1}$.}}

\tx For moderate and high mass accretion rates, $\tac \simless \tc$, the
cluster core does not revirialise immediately, and significant amounts
of mass are added before the core can adjust.  In order to quantify
how this changes the evolution of $\Rc$, we performed several direct
nbody simulations (using a tree-code derived from the SPH/N-body code
of Bate, Bonnell \& Price~1995) of clusters of 100 stars within a
radius of 0.1 pc undergoing constant accretion. These simulations were
compared with a toy model of the cluster core's evolution under the
influence of accretion. This model calculates, first, how the core
shrinks as the total energy decreases under accretion, assuming the
mass is added at a constant core radius (equation~4). The amount of
mass added is equal to the accretion rate multiplied by the timescale for
the process (the revirialisation timescale).  Secondly, the cluster
then revirialises at a radius corresponding to the new, lower, total
energy (equation~5). The timescale for the revirialisation is a free
parameter, on the order of a crossing time. By comparing the N-body
simulation with this toy model, we determined that the cluster can
effectively revirialise every $0.75\ \tc$ (Figure~1). In both cases,
the accreted gas is assumed to come from outside the core such that
$\Mc = \Ms$ (no unaccreted gas in core).  The two cases are well
matched after $\approx 1 \tc$, showing that the simple model for the
cluster's evolution under mass accretion is a good representation of
the dynamics involved.  Note that once the cluster has shrunk, the
crossing time is much less than the accretion timescale and the
variation in cluster radius follows equation~8.  It is also worth
noting that although the velocity dispersion initially decreases under
accretion (due to the conservation of momentum), the subsequent
revirialisation will increase the velocity dispersion as potential
energy is transferred into kinetic energy (equation~6). For isolated
clusters, the comparisons are only valid for approximately a
relaxation time (several crossing times for 100 stars), as an isolated
cluster of 100 stars starts to dissolve (due to the formation of a
hard binary which can absorb the total energy of the cluster).
However, if the 100 stars are embedded in a much larger system, it
does not dissolve, but simply exchanges members with the surrounding
cluster.

Thus, for most accretion rates, the rate at which the cluster (and
core) shrinks depends not only on the accretion itself but also on the
crossing time. This is important as it decreases the differences
between the rates of shrinkage (equation~8,~9 and~10) for the various
gas dynamics scenarios. The crossing time depends on the amount of
mass in the cluster core, such that the crossing time is significantly
shorter when all the accreted gas is initially contained within the
core as the velocity dispersion is larger. Thus, even though the core
does not shrink as much with accreted mass in this case (equation~9),
the fact that it can adapt much more quickly balances this difference and
the core actually evolves in a similar manner. In the following
sections we assume that the accreted gas primarily comes from outside
the cluster such that $\Mc \propto \Ms$.

\section{Timescales: Collisions versus accretion } 

\tx In order to ascertain how and when collisions become important,
we need to quantify and compare the timescales for collisions 
and accretion. The accretion timescale, \tacc, defined as the time it
takes to double a star's mass is simply given by
$$\tac = {\ms \over \msdot},\eqno(11)$$ where $\ms$ is the star's mass and
$\msdot$ is the stellar mass accretion rate.  The collisional
timescale has a more complicated form (see Binney \& Tremaine~1987 for
a derivation), depending on the stellar velocity dispersion, \vdisp,
the stellar density $n$, and the radius at which mergers occur, $\rs$:
$$ {1 \over \tcol} = 16 \sqrt{\pi} n \vdsp \rs^2 \biggl(1 +{G \ms
\over 2 \vdsp^2 \rs}\biggr).\eqno(12)$$ The last term incorporates the effects
of gravitational focusing in the cluster. Gravitational focussing has
an important effect on the collision rate for the cases considered here.

\vbox{
\figure{2}{S}{0mm}{\vskip-0.25truein\centerline{\vbox{\psfig{figure=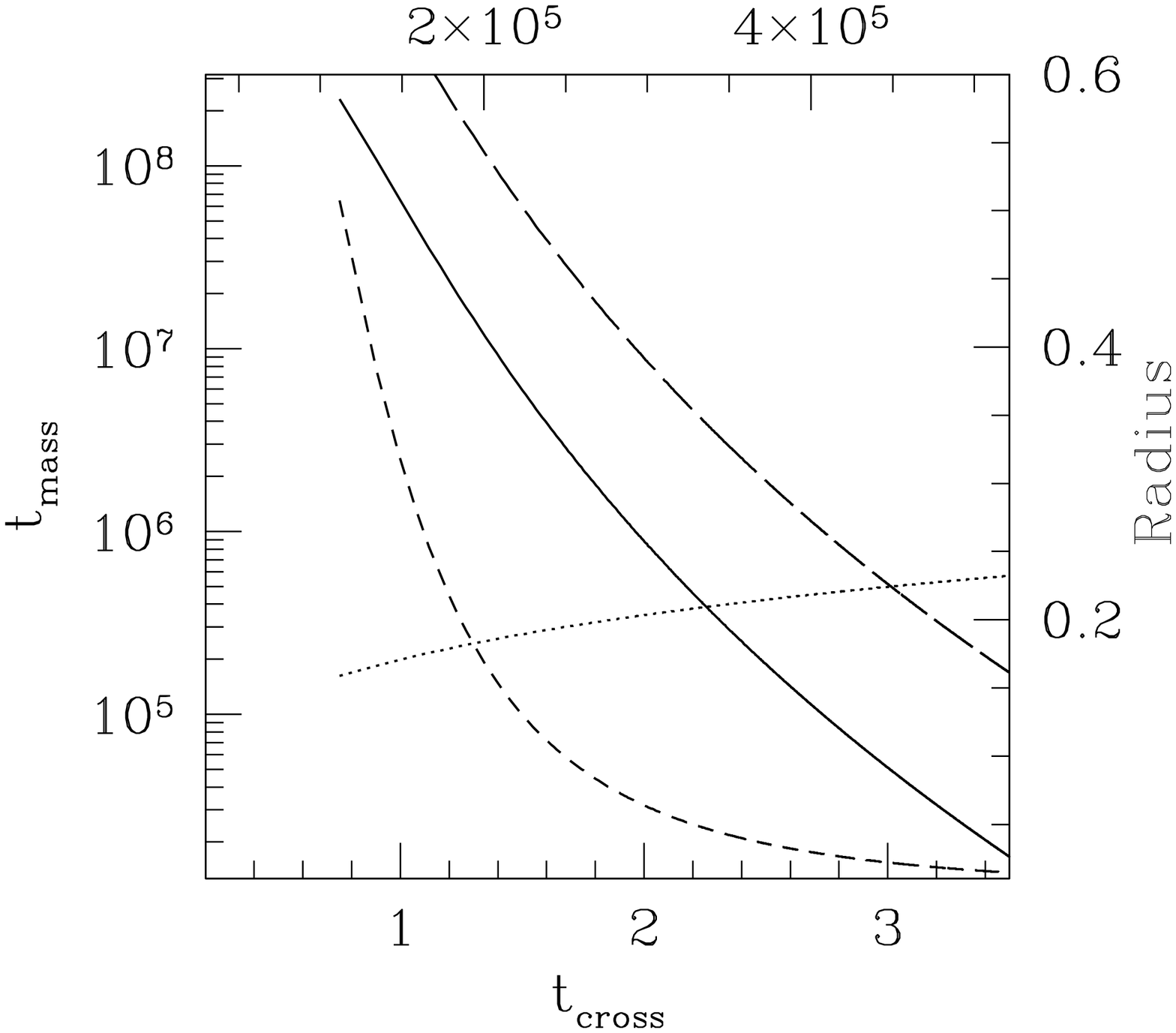,width=3.truein,height=3.truein,rwidth=3.truein,rheight=3.truein}}} 
\break\noindent
{\bf Figure 2.}  The timescale to double a star's mass either by
collisions (solid and long-dashed lines) or through accretion 
(dotted line) is plotted versus time in units of the initial
crossing time, $\tc = 1.5 \times 10^5$ years, (and in years) for a
cluster of 100 stars with an initial radius of 0.1 pc.
The accretion timescale assumes an accretion rate of $2 \times 10^{-6}$
\solmas yr$^{-1}$. The merger radius for collisions is taken to be $0.1$ 
(solid line) and $0.01$ au (long-dashed line).  The cluster radius in units of 
the initial radius is also plotted (short-dashed line).}}

For collisions to be at all relevant, the cluster core must be
sufficiently dense. Accretion onto the core of a cluster decreases its
radius and increases its density  until collisions eventually become
significant. Figure~2 illustrates the evolution of the accretion and
collisional timescales for a cluster of 100 stars initially contained
within a radius of 0.1 pc. The cluster members are all initially 0.1
\solmas\ and accrete at a rate of $2 \times 10^{-6} \solm {\rm yr}^{-1}$. The
collisional timescale is calculated assuming stellar merger radii of $0.1$ au
and $0.01$ au ($\rs \approx 21.5$ and $2.15$ \solrad).  We assume that
$\Mc \propto \Ms$ and neglect the mass of any gas in the core. The
main effects of a non-zero gas mass are to decrease the rate of shrinkage
with accretion and to decrease the crossing time. The net effect
is little different from Figure~2.

Figure~2 illustrates the relevant timescales for doubling a star's
mass ($t_{\rm mass}$).  The accretion timescale remains approximately
constant at a few $\times 10^5$ years, while the collision timescale
decreases from $>10^8$ years to $<10^5$ years.  Note that the initial
cluster crossing time is $\tc = 1.5 \times 10^5$ years and that in
both cases, the \tcoll\ becomes smaller than \tacc\ in $< 3 \tc$, or
$< 4.5 \times 10^5$ years. Thus, collisions become the dominant mode of
increasing a star's mass in the cluster core within $5 \times 10^5$ years.

The question of what the stellar merger radius should be is not easy
to answer as the protostars considered are still in their
pre-main-sequence phase, accreting from the surrounding material and
their accretion discs. They are in general significantly larger and
fluffier than their main sequence counterparts.  Pre-main sequence
stars, of various masses, less than $10^6$ years old, most probably
have radii in the range 5 to 20 \solrad\ (Palla \& Stahler~1993; Beech
\& Mitalas~1994).  Furthermore, the presence of an accreting envelope
and disc aids in the capture and merger process (eg Clarke \&
Pringle~1993; Hall, Clarke \& Pringle~1996), thus giving the stars an
effectively larger radius.  For the remainder of this paper, we assume
a merger radius of 0.1 au (21.5 \solrad).  It is worth noting that a
smaller merger radius still allows for the collisional buildup of
massive stars. However, it does imply a larger mean stellar mass at
the time when collisions become important as it allows for more
accretion.

\section{Collisions and massive stars} 

\vbox{
\figure{3}{D}{0mm}{\vskip-0.9truein\centerline{\vbox{\psfig{figure=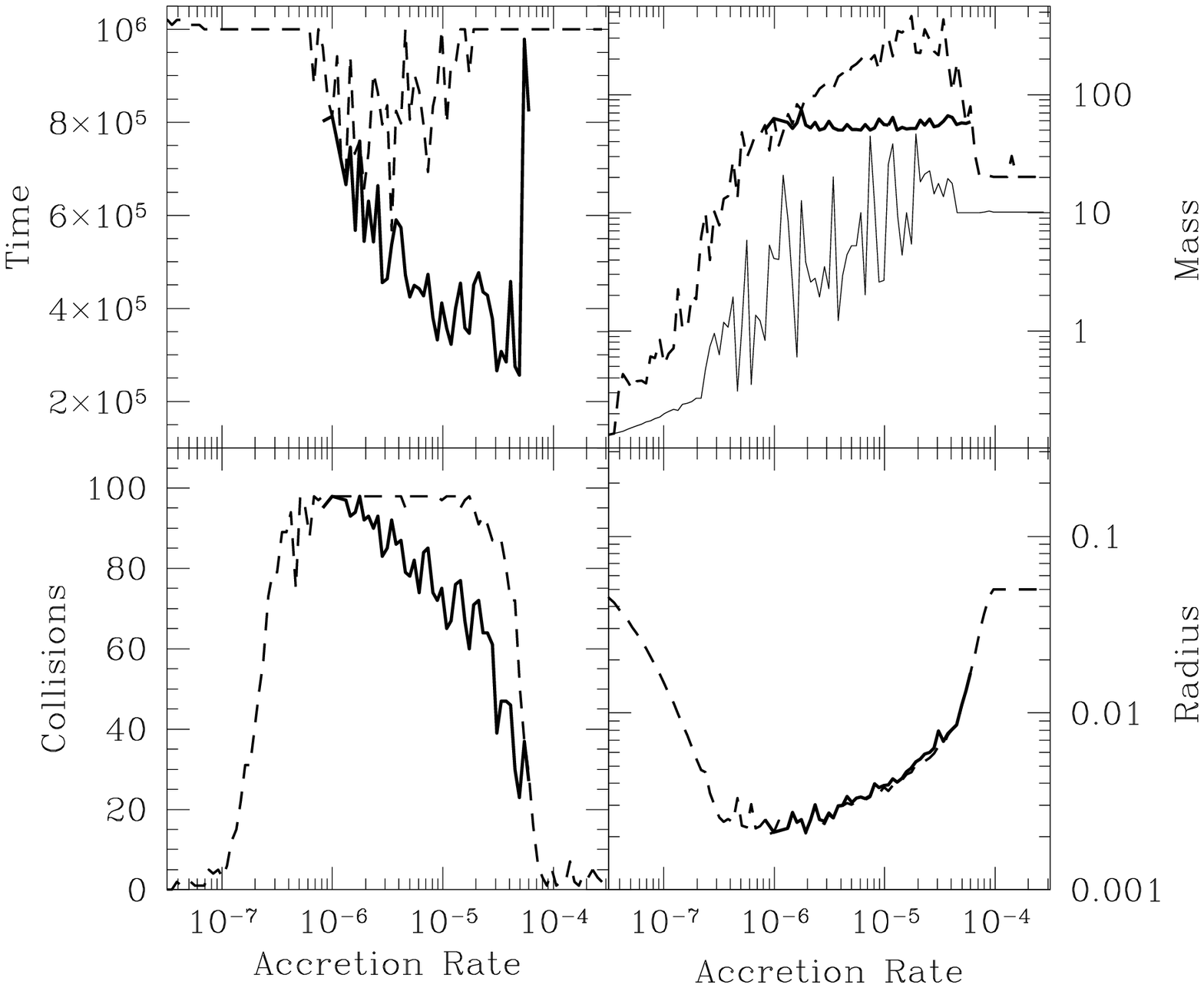,width=6.truein,height=6truein,rwidth=5.5truein,rheight=5.5truein}}} 
\break\noindent
{\bf Figure 3.} The four panels represent the toy-model calculations
of the central core of a cluster containing 100 stars initially within
0.1 pc, evolving under mean stellar accretion rates from $10^{-7} {\rm
to} 10^{-4}$
\solmas yr$^{-1}$. The stellar merger radius is taken to be 0.1 au. 
The top left panel shows the time, in years, at the end (dashed line)
of the evolution (either $10^6$ years or when only 2 stars are left),
and the time when a 50 \solmas\ star is first formed (heavy solid line). 
The top-right panel shows the maximum stellar mass (in \solmas) for
the same cases plus the median stellar mass at the end of the
evolution (light solid line). The bottom left panel shows the
number of collisions that occur for the two cases while the radius
(in pc) of the cluster core is shown in the bottom right panel.}}

\tx In the previous sections, we have seen how the central cores of stellar
clusters undergoing accretion eventually shrink sufficiently that
collisions dominate the increase in stellar masses. In this section we
use the same  toy model described above but modified to include 
collisions and competitive accretion (Zinnecker~1982).  Accretion is
allowed onto all stars in proportion to their contribution to the
gravitational potential, ie
$$\msdot = {\ms^2 \over \Sigma_i {{\ms}_i}^2} \mcdot.\eqno(13)$$ This
is equivalent to each star accreting with a Bondi-Hoyle type accretion
rate (eg Ruffert~1996) assuming that the accreting material is at a
uniform density and each star is moving through the medium with the
same velocity.  The collisions occur on the collisional timescale of
each star,
\tcoll\ (equation 12), and the stars are assumed to merge when they
collide.  The collisions are assumed to be 100 \% efficient in that
no mass is lost. In reality, some small amounts of mass will be
lost during collisions (eg Davies \etal~1993). Mass loss rates
of up to 25 \% of the stellar masses were attempted with no
significant differences in the results. 
Accretion is not allowed onto stars more massive than 10
\solmas\ so that stars more massive than this need to form through
collisions.  The cluster evolution is followed, allowing for
revirialisation every $0.75 \tc$, and  recalculating \tcoll\ for
each star.  

The number of stars is only affected by collisions. No stars are
allowed to leave or enter the cluster core. In reality some
interchange between the core and the surrounding cluster is expected,
introducing some lower-mass stars and removing some of the high-mass
products. Ejection of stars from the core require an interaction with
a hard binary in order to give a kick velocity greater than the
velocity dispersion.  The finite and, including circumstellar
material, relatively large size of the young stars limits such
ejections as the interaction will often lead to a merged object
(Davies, Benz \& Hills~1994).  Furthermore, the presence of the gas in
the cluster should decrease such ejections and, in general, it would
be preferentially the less massive stars that are lost from the core.
If many interchanges occur, then the introduction of low-mass stars
will increase the number of collisions required to form a massive
star.  In the extreme case, where many more stars leave the core then
enter it, an initially higher number of stars in the core may be
necessary. These additional complications are beyond the scope of the
present study but should not overtly change the results.

The models are followed for $10^6$ years or until only two stars
remain (a rare occurrence).  Models with varying numbers of stars in
the cluster core and varying core radii were investigated. In general,
cluster cores with \tcr\ significantly less than $10^6$ years (the
approximate age at which a complete mass spectrum is observed in the
ONC) are the ones in which significant collisional buildup of massive
stars occurs.

\vbox{
\figure{4}{D}{0mm}{\vskip-0.9truein\centerline{\vbox{\psfig{figure=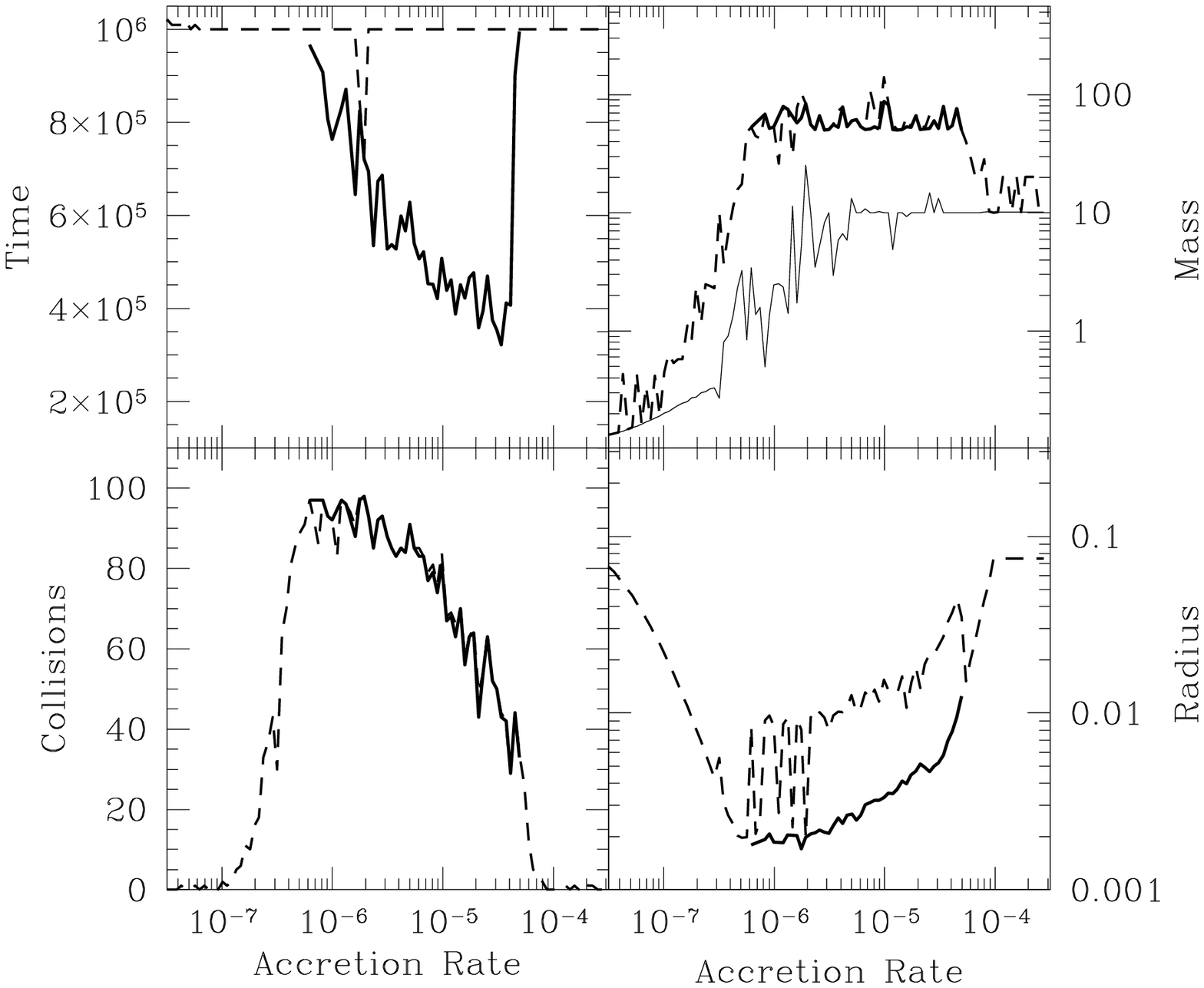,width=6.truein,height=6.truein,rwidth=5.5truein,rheight=5.5truein}}} 
\break\noindent
{\bf Figure 4.} Same as Figure~3 except for a cluster core of 100
stars initially within 0.15 pc and with a gas fraction of 50 \%
contributing to the gravitational potential. The gas is removed once a
star with $M \simgreat 50 \solm$ has formed. The removal occurs over
several crossing times.}}

\vbox{
\figure{5}{D}{0mm}{\vskip-0.9truein\centerline{\vbox{\psfig{figure=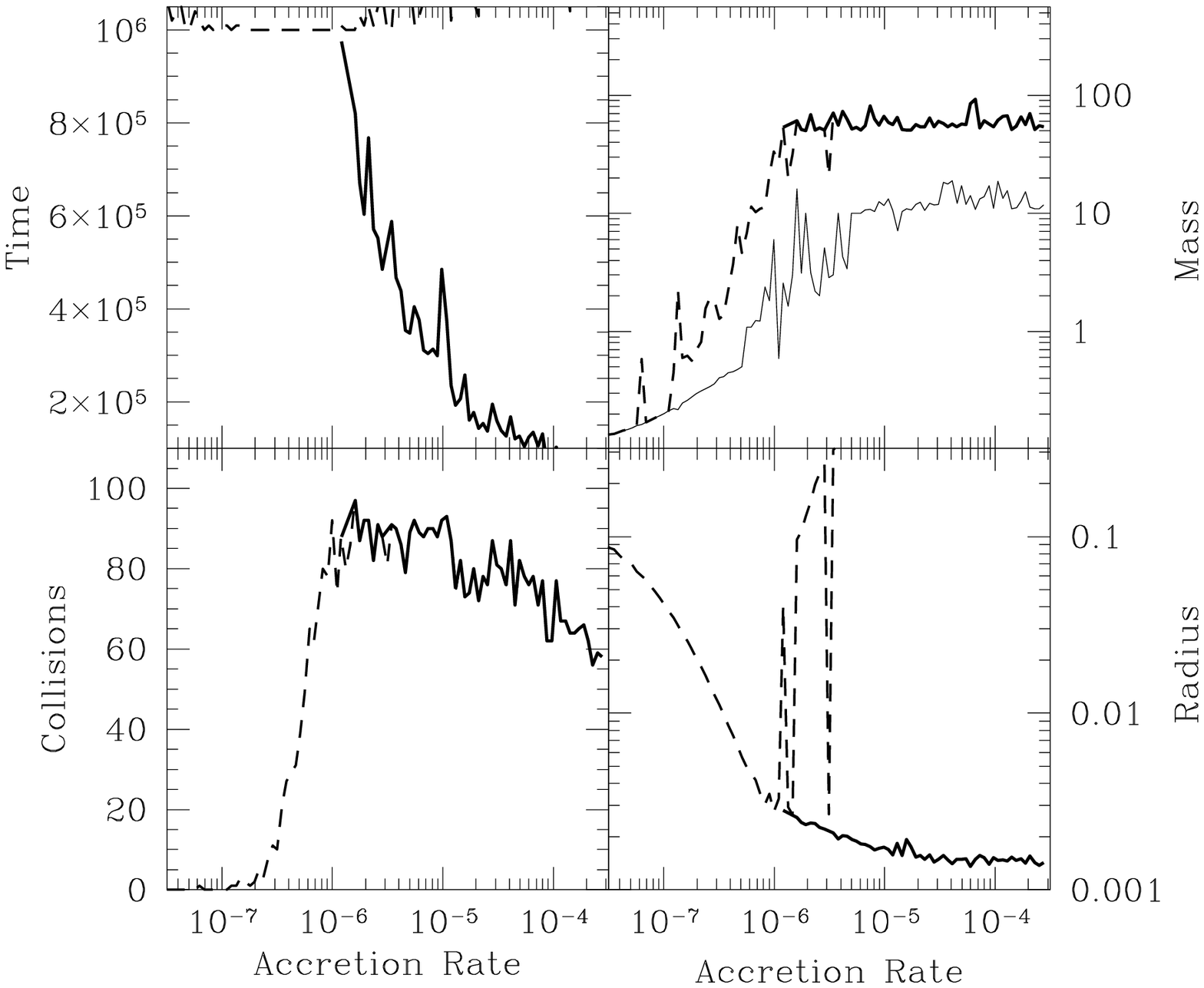,width=6.truein,height=6.truein,rwidth=5.5truein,rheight=5.5truein}}} 
\break\noindent
{\bf Figure 5.} Same as Figure~3 except for a cluster core of 100
stars initially within 0.15 pc and with a constant core mass. The gas
is initially contained within the core and has mass equal to that
required for the accretion to be sustained for a period of $2 \times
10^6$ years. The gas is removed once a star with $M \simgreat 50
\solm$ has formed. The removal occurs over several crossing times.}}

An illustrative case of a cluster core with 100 stars in 0.1 pc,
initial stellar masses of $0.1 \solm$ and a stellar merger radius of 0.1 au,
evolving under various mean stellar accretion rates is shown in Figure~3. This
figure shows the effects of accretion-induced collisions in forming
massive stars.  The top-left panel shows the period of time required
to form a star of 50 \solmas, and the time at which only 2
remain of the original 100 protostars (or $10^6$ years). It is worth
noting that the evolution requires several $\times 10^5$ to $10^6$
years to form a 50 \solmas\ star.  Thus, massive stars in these
clusters are expected to be considerably younger than the mean stellar
age, and in some cases, could be up to $10^6$ years
younger than the oldest low-mass stars.

The bottom-left panel of Fig.~3 shows the number of collisions that
occur and the bottom-right panel shows the cluster's core radius for
the same cases.  For small mean stellar accretion rates, few
collisions occur as the cluster does not shrink sufficiently within
$10^6$ years.  For large accretion rates, the mass is added very
quickly, $\tac << \tc$, and by the time the cluster revirialises, the
mean stellar mass is greater than 10 \solmas, halting the
accretion. In this case the cluster core does not shrink very much 
(half the initial core radius) and
hence collisions are rare. Maximal shrinkage occurs when $\tac
\approx
\tc \approx 10^6$ years. Once collisions dominate, the cluster evolves
due to the collisions and has no time to shrink further. This occurs
at slightly larger radii for larger accretion rates and the mean
stellar mass is then larger.

The top-right panel of Fig.~3 shows the maximum mass obtained at
either $10^6$ years, or when only 2 stars are left, or when a
collision has formed a star of mass $\ge 50 $ \solmas.  It is clear
from this panel that the maximum mass depends on both the accretion
rate and the number of collisions. The maximum mass is greater than 50
\solmas\ for accretion rates between $5 \times 10^{-7}$ and $5 \times
10^{-5}$ \solmas\ yr$^{-1}$. The median mass at the time that a 50
\solmas\ star is first formed is also shown in Fig.~3.  The median
mass typically lies between a few and 10 \solmas.  It is worth noting
that for accretion rates $> 10^{-5}$ \solmas\ yr$^{-1}$, accretion
alone will produce a 10 \solmas\ star in 10$^6$ years.

As noted above, the exact value of the stellar merger radius does not
have a large affect on the results. The above calculations were also
performed with a stellar merger radius of 5 \solrad\ and gave qualitatively
similar results, the main differences being the amount the cluster shrinks
as the smaller stellar merger radius necessitates a higher stellar density for 
collisions, and the mean stellar mass which is larger in the
smaller stellar merger radius case.

The maximum mass in Fig.~3 sometimes reaches well in excess of 100
\solmas.  This is clearly not physical and is due to the artificial
termination points for the evolution (2 stars left, or $10^6$
years). A more realistic termination point requires some process that
stops the collisions. Up to now, we have ignored the
contribution to the potential of the gas being accreted. The gas
component in young clusters typically comprises 50 to 90 \% of the
total mass in the cluster (Lada \etal~1993). The exception to this is
the Trapezium where the most massive star, $\thetaonec$, has
presumably removed all the gas through its ionising photons and wind. This
offers an appealing process of halting further collisions.

In order to include the effect of the gas on the potential, we
performed several calculations with either a gas contribution equal to
that of the stars or a constant core mass and thus an ever decreasing
gas mass.  In both cases, the gas is presumed to be removed over
several crossing times once a star of sufficient mass (50 \solmas\ in
this example) is formed.  The gas is expelled due to the ionising
photons and wind from the massive star.

Firstly, we consider the effect of the gas
when it contributes equally with the stellar mass.  Figure~4 shows the
evolution in this case for a cluster core of 100 stars initially
within a radius of 0.15 pc (the increased radius balances the added
mass component of the gas, so that the crossing time is similar). The
evolution is very similar to the previous one shown in Figure~3 until
a 50 \solmas star is formed.  At this point, the gas is removed and
the cluster relaxes, increasing its radius, to account for the loss of
potential energy. The larger radius, and hence smaller stellar
density, effectively eliminates any further collisions. This impedes
the formation of the very massive stars ($M>100 \solm$).

Alternatively, if the gas is completely contained within the core
initially, and the mass of the core is constant, then the core radius
does not shrink as rapidly with accreted mass ($\Rc \propto
\Ms^{-2}$), but this is somewhat offset by the much shorter crossing
time as the velocity dispersion is higher due to the larger core mass.
Such an evolution is illustrated in Figure~5.  The initial gas mass is
set to twice that required for the desired accretion rate over $10^6$
years.  Thus, the star formation efficiency increases from near zero
to a maximum of 50 \%.  The minimum cluster radius is well represented
by equation~9: $\Rc / R_{\rm init} = (\msdot \times t / {\ms}_{\rm
init})^{-2}$.  The main difference between the cluster evolution in
this case and in the previous ones is that, although the cluster only
shrinks as $\Ms^{-2}$, due to the large amounts of gas included in the
cluster core, the initial crossing time is much shorter. Thus, it can
adapt very quickly and there is no maximum value of the mean stellar
accretion rate for which significant shrinkage occurs. Any large rate
($\msdot \simgreat 10^{-6}$ \solmas yr$^{-1}$) is sufficient to drive
the collisional buildup of massive stars.  The large amounts of gas
included in the cluster core result in the core being unbound and
expanding once the gas is removed by a newly formed massive star ($\ms
\simgreat 50 \solm$). In this case, the massive stars quickly expand
into the rest of the cluster.

\section{Binary systems} 

\tx So far, we have neglected binary systems and their possible role
in the evolution of the cluster core and in the formation of
massive stars. Binary systems are commonly employed as an
energy source in globular clusters to stave off core-collapse.
The binaries inject energy through binary-single interactions
transforming their potential energy into the kinetic energy
of the cluster.

In the scenario considered here, the cluster is shrinking due to the
accretion. The rate of shrinkage depends on the relation between the
accretion timescale and the crossing time. In the extreme case, $\tac >>\tc$,
the shrinkage goes as $\Rc \propto \Ms^{-3}$. This is the same rate as
is found for a binary system that is accreting zero-angular
momentum matter (Smith, Bonnell \& Bate~1997). Thus, unless a primordial
binary is initially capable of affecting the cluster dynamics, the subsequent
accretion and shrinkage should not make it an important energy source for
the cluster core.

We do know, however, that many massive stars are in close binary
systems (Garmany, Conti \& Massey~1980, Mason \etal~1996) so that any
theory for their formation should be able to explain why they are in
such systems. Their multiplicity can easily be explained in the
context of the model presented here where massive star formation is
due to collisions. For collisions to be important, many more
interactions will occur at slightly greater periastron separations.
These interactions can result in the formation of binary systems
through tidal capture (eg Mardling~1996). Tidal capture can produce
binaries at a few stellar radii but will not form very wide
systems. Furthermore, the next interaction can result in a loss of
kinetic energy from the binary and thus help the stellar merger
rate. In any case, the end result is likely to be of a binary system
comprising two of the remaining (proto)stars. The resultant binary is
likely to comprise two relatively massive stars, as they both should
have accreted a significant amount of mass, and/or undergone
collisional mass buildup, before capture. The number of binary systems
should be large amongst the massive stars because the tidal capture
rate must exceed that for collisions.

Lastly, binary systems among the massive stars can provide a test of
this mechanism for the formation of massive stars.  The maximum
separation of such systems should not, in general, be larger than the
hard/soft boundary typical for the stellar density when the collisions
occurred. Except for the possibility of binary formation at the very end of the
collisional buildup of massive stars, all binaries of separations
greater than this critical, density-dependent, separation are broken
up by encounters within the core.

\section{Implications} 

\tx There are several implications that can be gleaned from the models
presented here.  The accretion-induced collisions mechanism requires
that the massive stars form in the centre of a large stellar cluster.
Thus massive stars should never form in isolation, and those that are
found in isolation need to have been ejected from such dense systems.

Accretion-induced collisions
can, under reasonable accretion rates, form massive stars in less than
$10^6$ years. The process does require relatively dense initial
conditions, in order to be able to accrete and shrink sufficiently quickly
so that collisions are frequent. This is equivalent to requiring
$\tc \approx \tac << t_{\rm end}$ where $t_{\rm end} \approx 10^6$ years
or some similar maximum time. For initial masses $\approx 0.1 \solm$, this
implies minimum stellar densities of $\simgreat 10^4$ stars pc$^{-3}$.
Such densities are typical only of the central regions of large
clusters such as the Orion Nebula cluster, where these massive stars
are indeed found (McCaughrean \& Stauffer~1994; Hillenbrand~1997).

The time required for accretion-induced collisions to form
a massive star may imply a considerable age difference between the
older low-mass stars and the younger high-mass stars. This
age difference is at least several $\times 10^5$ years 
and can be as much as $10^6$ years, depending on the
initial cluster density and the accretion rate.
Thus stars such as $\thetaonec$ in the
ONC may have only recently ``turned on'' and started to affect the gas
in the cluster and in discs around the low-mass stars.
The Trapezium is then the remnant of the central dense core where
the collisions occurred and which has recently expanded to its present
configuration due to the gas expulsion.

The collisional buildup of massive stars also has implications for the
intermediate-mass stars. It is these stars that, having attained their
mass through accretion in the cluster core, are required to collide to
form the massive stars. Thus, the formation of the massive stars
implies a removal of intermediate mass stars. Systems with massive
stars could then demonstrate a relative lack of intermediate-mass
stars.  There is a tantalising hint of the relative paucity of
intermediate-mass stars in the ONC (Hillenbrand~1997) and in a few
other clusters (Wilner \& Lada~1991) although it could be due simply
to statistical fluctuations.

Finally, since, in this model, massive ($m \simgreat 10 \solm$) and
intermediate mass ($m \simless 10 \solm$) stars are formed by
different mechanisms, the resultant mass distribution may display some
structure at $\approx 10 \solm$ (not a single power-law IMF).

\section{Conclusions} 

\tx Accretion-induced collisions in the core of a dense, young stellar cluster
can form massive stars. This circumvents the problem of accreting
directly onto stars with masses $> 10 \solm$ where radiation pressure
can halt the infall. Gas accretion onto individual members of the core
of a stellar cluster forces the core to shrink. If the accretion
timescale is comparable to the initial crossing time, the core can
shrink sufficiently that collisions become significant. The collisions
involve intermediate mass stars that are formed by accretion onto
initially lower mass protostars. The actual stellar merger radius does not
overtly affect this process as the cluster core will shrink due to the
accretion until collisions become dominant.  The collisions are most
probably halted by the ejection of the gas contained in the core once
a sufficiently massive star is formed ($\approx 50 \solm$).

This mechanism for forming massive stars implies that they need to
form in the centre of rich, dense young stellar clusters. They could
appear significantly younger than the mean age of the low-mass stars.
There is also the potential that the collisions will deplete the
number of intermediate mass stars as it is they that will collide to
form the massive stars. Close binary systems should be common amongst
the massive stars, formed through tidal capture of stars which have
close interactions but do not collide.

\section*{Acknowledgments}

\tx We thank Cathie Clarke, Jim Pringle, Melvyn Davies, Hal Yorke 
and Russell Cannon for many insightful discussions.  We also thank an
anonymous referee whose suggestions helped significantly improve the
text.  IAB acknowledges support from a PPARC advanced fellowship.

\section*{References}

\bibitem Bastien P., 1981, A\&A, 93 160

\bibitem Bate M.R., Bonnell I.A., Price N.M, 1995, MNRAS, 277, 362

\bibitem Beech M., Mitalas R., 1994, ApJS, 95, 517

\bibitem Binney J., Tremaine S., 1987, in Galactic Dynamics, Princeton 
University.

\bibitem Bonnell I. A., 1994, MNRAS, 269, 837

\bibitem Bonnell I. A., Bate M. R., Clarke C. J., Pringle J. E, 1997, MNRAS, 
285, 201

\bibitem Bonnell I. A., Davies M. B., 1997, MNRAS, in press

\bibitem Clarke C. J., Pringle J. E., 1992, MNRAS, 255, 432

\bibitem Clarke C. J., Pringle J. E., 1993, MNRAS, 261, 190 

\bibitem Davies M.B., Benz W.,  Hills J.G., 1994, ApJ, 424, 870

\bibitem Davies M. B., Ruffert M., Benz W., M\"uller E., 1993, A\&A, 272, 430

\bibitem Garmany C. D., Conti P. S, Massey P., 1980, ApJ, 242, 1063

\bibitem Hall, S. M., Clarke C. J., Pringle J, E., 1996, MNRAS, 278, 303

\bibitem Hillenbrand L. A., 1997, AJ, 113, 1733

\bibitem Hillenbrand L. A., Hartmann L., 1997, preprint

\bibitem Lada E. A., Strom K. M., Myers P. C., 1993, in {\sl
Protostars and Planets III}, eds E. Levy and J. Lunine, University of
Arizona, p.  245

\bibitem Larson R. B., 1992, MNRAS, 256, 641

\bibitem Laughlin G., Bodenheimer P., 1994, ApJ, 436, 335

\bibitem Mardling R. A., 1996, in {\sl Evolutionary Processes in Binary Stars},
eds R. A. M. J. Wijers, M. B. Davies and C. A. Tout, Kluwer, Dordrecht, p. 81

\bibitem Mason B. D., Hartkopf W. I., Gies D. R., McAlister H. A., Bagnuolo 
W. G., 1996, in {\sl The Origins, Evolution and Destinies of Binary
Stars in Clusters}, eds E. F. Milone and J. C. Mermilliod, ASP, p. 40

\bibitem McCaughrean M. J., Stauffer J, R., 1994, AJ, 108, 1382

\bibitem Murray S. M., Lin D. N., 1996, ApJ, 467, 728

\bibitem Palla F., Stahler S. W., 1993, ApJ, 418, 414

\bibitem Price N. M., Podsiadlowski Ph., 1995, MNRAS, 271 1041

\bibitem Ruffert M., 1996, A\&A, 311, 817

\bibitem Smith K. W., Bonnell I. A., Bate M. R., 1997, MNRAS, 288, 1041

\bibitem Wilner D. J., Lada C. J., 1991, AJ, 102, 1050
 
\bibitem Wolfire M. G., Cassinelli J. P., 1987, ApJ, 319, 850

\bibitem Yorke H. W. 1993, in {\sl Massive Stars: Their Lives in the 
	interstellar Medium}, eds. J. Cassinelli, E. Churchwell, 
	ASP Conf. Ser. 35:45--55.

\bibitem Yorke H. W., Bodenheimer P., Laughlin G., 1995, ApJ, 443, 199

\bibitem Yorke H. W., Kr\"ugel E., 1977, A\&A, 54, 183

\bibitem Yorke H. W., Welz A., 1996, A\&A, 315, 555

\bibitem Zinnecker H., 1982 in {\sl Symposium on the Orion Nebula to
Honour Henry Draper}, eds A. E. Glassgold \etal, New York Academy of
Sciences, p. 226

\bibitem Zinnecker H., McCaughrean M. J., Wilking B. A., 1993 in {\sl
Protostars and Planets III}, eds E. Levy and J. Lunine, University of
Arizona, p. 429

\vfill\eject\end